\documentclass[openany]{nature}
\usepackage[left]{lineno}
\usepackage{graphicx}
\usepackage{float}
\usepackage{amsmath}
\usepackage{amssymb}
\usepackage{subfigure}
\usepackage{cases}
\usepackage{mathrsfs}
\usepackage[flushleft]{threeparttable}
\usepackage{amssymb}
\usepackage{pifont}
\usepackage{epstopdf}
\usepackage{xcolor}
\usepackage{hyperref}
\usepackage[all]{hypcap}
\usepackage{mwe,tikz}
\usepackage{bm}
\usepackage{multirow}
\usepackage{longtable}
\usepackage{xspace}
\usepackage{rotating}
\usepackage{CJK}
\usepackage{url}
\usepackage{upgreek}
\usepackage{booktabs}
\usepackage{textcomp}
\usepackage{makecell}
\usepackage{enumitem}
\usepackage{csquotes}
\usepackage{soul}
\usepackage{tabularx}
\usepackage{overpic}
\usepackage{caption}
\usepackage{float}
\usepackage{threeparttable}
\usepackage{threeparttablex}

\makeatletter
\renewcommand{\baselinestretch}{1.2} 
\def\emph#1 {\textit{ #1 } }

\let\saved@includegraphics\includegraphics
\AtBeginDocument{\let\includegraphics\saved@includegraphics}
\renewenvironment*{figure}{\@float{figure}}{\end@float}
\makeatother

\newcommand{\apj}{Astrophys. J.}

\newcommand{\pasp}{Publ. Astron. Soc. Pac.}
\newcommand{\apjs}{Astrophys. J. Supp.}
\newcommand{\araa}{Annu. Rev. Astron. Astrophys.}
\newcommand{\mnras}{Mon. Not. R. Astron. Soc.}
\newcommand{\apjl}{Astrophys. J. Let.}
\newcommand{\aap}{Astron. Astrophys.}
\newcommand{\aj}{Astron. J.}
\newcommand{\nat}{Nature}

\newcommand{\pasj}{Publ. Astron. Soc. Jpn.}

\newcommand{\ssr}{Space. Sci. Reviews.}

\title{Soft X-ray prompt emission from a high-redshift gamma-ray burst EP240315a}

\author{Y. Liu$^{1}$\thanks{These authors contributed equally to this work}, H. Sun$^{1*}$, D. Xu$^{1*}$,
D.~S. Svinkin$^{2}$, J. Delaunay$^{3}$, N. R. Tanvir$^{4}$,
H. Gao$^{5,6}$\thanks{E-mail: gaohe@bnu.edu.cn},
C. Zhang$^{1}$\thanks{E-mail: chzhang@bao.ac.cn},
Y. Chen$^{7}$\thanks{E-mail: ychen@ihep.ac.cn},
X.-F. Wu$^{8}$\thanks{E-mail: xfwu@pmo.ac.cn}, B. Zhang$^{9,10}$
% ...
}

\begin{document}
\captionsetup[table]{name={\bf Table}}
\captionsetup[figure]{name={\bf Fig.}}

\maketitle

\begin{affiliations}
\item{National Astronomical Observatories, Chinese Academy of Sciences, Beijing 100101, China.}
\item{Ioffe Institute, Politekhnicheskaya 26, 194021 St. Petersburg, Russia.}
\item{Department of Astronomy and Astrophysics, The Pennsylvania State University, 525 Davey Lab, University Park, PA 16802, USA.}
\item{School of Physics and Astronomy, University of Leicester, LE1 7RH, UK.}
\item{Institute for Frontier in Astronomy and Astrophysics, Beijing Normal University, Beijing 102206, China.}
\item{Department of Astronomy, Beijing Normal University, Beijing 100875, China.}
\item{Key Laboratory of Particle Astrophysics, Institute of High Energy Physics, Chinese Academy of Sciences, Beijing 100049, China.}
\item{Purple Mountain Observatory, Chinese Academy of Sciences, Nanjing 210023, China.}
\item{Nevada Center for Astrophysics, University of Nevada Las Vegas, NV 89154, USA.}
\item{Department of Physics and Astronomy, University of Nevada Las Vegas, NV 89154, USA.}
\item{School of Astronomy and Space Science, University of Chinese Academy of Sciences, Chinese Academy of Sciences, Beijing 100049, China.}
\item{INAF - Istituto di Astrofisica e Planetologia Spaziali, via Fosso del Cavaliere 100, 00133 Rome, Italy.}
\item{INAF - Osservatorio Astronomico di Brera, Via E Bianchi 46, I-23807 Merate (LC), Italy.}
\item{INFN – Sezione di Milano-Bicocca, piazza della Scienza 3, I-20126 Milano (MI), Italy.}
\item{Cosmic Dawn Center (DAWN), Copenhagen 2200, Denmark.}
\item{Niels Bohr Institute, University of Copenhagen, Copenhagen 2200, Denmark.}
\item{Department of Astrophysics/IMAPP, Radboud University Nijmegen, Nijmegen, 6500 GL, The Netherlands.}
\item{University of Rome Tor Vergata, Department of Physics, via Cracovia 50, 00100 Rome, Italy.}
\item{INAF - via Parco del Mellini, 00100 Rome, Italy.}
\item{GEPI, Observatoire de Paris, Université PSL, CNRS, Meudon 92190, France.}
\item{School of Astronomy and Space Science, Nanjing University, Nanjing 210093, China.}
\item{Key Laboratory of Modern Astronomy and Astrophysics (Nanjing University), Ministry of Education, China.}
\item{Universit\'{e} de la C\^ote d'Azur, Observatoire de la C\^ote d'Azur, CNRS, Artemis, Nice F-06304, France.}
\item{Aix Marseille Univ, CNRS, LAM, Marseille, France.}
\item{ARC Centre of Excellence for Gravitational Wave Discovery (OzGrav), Hawthorn, VIC 3122, Australia.}
\item{Centre for Astrophysics and Supercomputing, Swinburne University of Technology, Hawthorn, Victoria, Australia.}
\item{Sydney Institute for Astronomy, School of Physics, The University of Sydney, NSW 2006, Australia.}
\item{European Space Agency, ESTEC, Keplerlaan 1, NL-2200 AG, Noordwijk, The Netherlands.}
\item{Department of Astronomy, School of Physics, Huazhong University of Science and Technology, Wuhan, Hubei 430074, China.}
\item{David A. Dunlap Department of Astronomy and Astrophysics, University of Toronto, 50 St. George Street, Toronto, ON M5S 3H4, Canada.}
\item{Dunlap Institute for Astronomy and Astrophysics, University of Toronto, 50 St. George Street, Toronto, ON M5S 3H4, Canada.}
\item{Racah Institute of Physics, The Hebrew University of Jerusalem, Jerusalem 91904, Israel.}
\item{Department of Physics and Astronomy, University of Sheffield, Sheffield S3 7RH, UK.}
\item{School of Physics and Centre for Space Research, University College Dublin, Dublin 4, Ireland.}
\item{Center for Space Science and Technology, University of Maryland Baltimore County, 1000 Hilltop Circle, Baltimore, MD 21250, USA.}
\item{Astrophysics Science Division, NASA Goddard Space Flight Center, Greenbelt, MD 20771, USA.}
\item{INAF - Osservatorio di Astrofisica e Scienza dello Spazio, Via Piero Gobetti 93/3, Bologna 40129, Italy.}
\item{Instituto de Astrof\'{i}sica de Canarias, iac, La Laguna, S/C de Tenerife, Spain.}
\item{Departamento de Astrof\'isica, Univ. de La Laguna, La Laguna, Tenerife, Spain.}
\item{Kavli Institute for Astrophysics and Space Research, Massachusetts Institute of Technology, Cambridge 02139, MA, USA.}
\item{School of Physics and Astronomy, Sun Yat-Sen University, Zhuhai 519082, China.}
\item{Department of Astronomy \& Astrophysics, University of Toronto, Toronto, ON M5S 3H4, Canada.}
\item{Dunlap Institute for Astronomy \& Astrophysics, University of Toronto, Toronto, ON M5S 3H4, Canada.}
\item{Department of Physics, University of Cagliari, SP Monserrato-Sestu, km 0.7, 09042 Monserrato, Italy.}
\item{National Space Science Center, Chinese Academy of Sciences, Beijing 100190, China.}
\item{Max-Planck-Institut für extraterrestrische Physik, Giessenbachstrasse 1, 85748 Garching, Germany.}
\item{Innovation Academy for Microsatellites, Chinese Academy of Sciences, Shanghai 201210, China.}
\item{Shanghai Institute of Technical Physics, Chinese Academy of Sciences, Shanghai 200083, China.}
\item{Key Laboratory of Technology on Space Energy Conversion, Technical Institute of Physics and Chemistry, CAS, Beijing 100190, China.}
\item{University of Chinese Academy of Sciences, Beijing 100049, China.}
\item{Department of Astronomy, School of Physical Sciences, University of Science and Technology of China, Hefei 230026, China.}
\item{Xinjiang Astronomical Observatory, Chinese Academy of Sciences, Urumqi, Xinjiang 830011, China.}
\item{North Night Vision Technology Co., LTD, Nanjing, China.}
\item{Guangxi Key Laboratory for Relativistic Astrophysics, School of Physical Science and Technology, Guangxi University, Nanning 530004, China.}
\item{Yunnan Observatories, Chinese Academy of Sciences, Kunming 650011, China.}
\item{Institute of Space Sciences (ICE), Consejo Superior de Investigaciones Científicas (CSIC), Barcelona, Spain.}
\item{Institut d’Estudis Espacials de Catalunya (IEEC), Barcelona, Spain.}
\item{Institute of Electrical Engineering, Chinese Academy of Sciences, Beijing 100190, China.}
\item{Department of Physics, Air University Islamabad, E-09 Sector PAF Complex 44000, Islamabad, Pakistan}
\item{Physics Department, Tsinghua University, Beijing, 100084, China.}
\item{Institute of Astrophysics, Central China Normal University, Wuhan 430079, China.}
\end{affiliations}

\begin{abstract}
Long gamma-ray bursts (GRBs) are believed to originate from core collapse of massive stars. High-redshift GRBs can probe the star formation and reionization history of the early universe, but their detection remains rare. Here we report the detection of a GRB triggered in the 0.5–4 keV band by the Wide-field X-ray Telescope (WXT) on board the Einstein Probe (EP) mission, designated as EP240315a, whose bright peak was also detected by the Swift Burst Alert Telescope and Konus-Wind through off-line analyses. At a redshift of $z=4.859$, EP240315a showed a much longer and more complicated light curve in the soft X-ray band than in gamma-rays. Benefiting from a large field-of-view ($\sim$3600 deg$^2$) and a high sensitivity, EP-WXT captured the earlier engine activation and extended late engine activity through a continuous detection. With a peak X-ray flux at the faint end of previously known high-$z$ GRBs, the detection of EP240315a demonstrates the great potential for EP to study the early universe via GRBs.
\end{abstract}


Gamma-ray bursts (GRBs) are explosive events that release extreme amounts of energy, with typical timescales of milli- to hundreds of seconds, that can be detected even when they occur in the very early Universe\cite{Zhang2018pgrb.book}. Long GRBs are believed to originate from the death of massive stars\cite{Galama1998, Hjorth2003Natur, Woosley2006ARA&A}, serving as a unique tool for studying the history of star formation at high redshifts\cite{Yuksel2008ApJ, Kistler2009ApJ} and the characteristics of the local environment during those epochs\cite{Fynbo2006A&A, Heintz2018MNRAS}. 
So far, the detection rate of the high-redshift GRBs by the gamma-ray detectors is notably low, with a mere 3\% or less of the entire Swift-triggered sample and approximately 10\% or fewer of the redshift-known sample having redshifts exceeding $z>4.5$\cite{Cusumano2006Natur, Salvaterra2009Natur, Cucchiara2011ApJ, Lien2016ApJ}.

\begin{figure}
\centering
\begin{tabular}{c}
\begin{overpic}[width=0.6\textwidth]{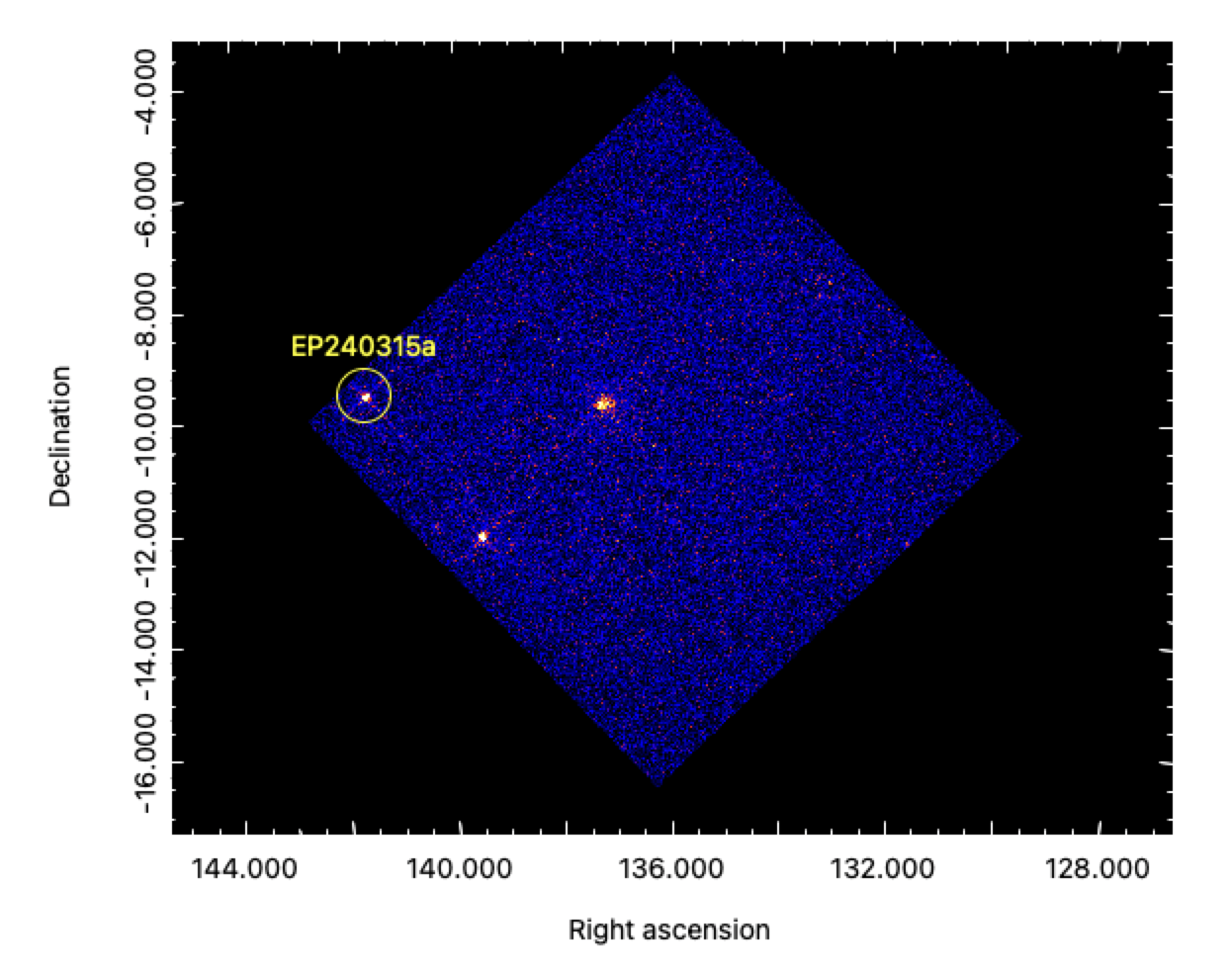}\put(0, 75){\bf a}\end{overpic} \\
\begin{overpic}[width=0.6\textwidth]{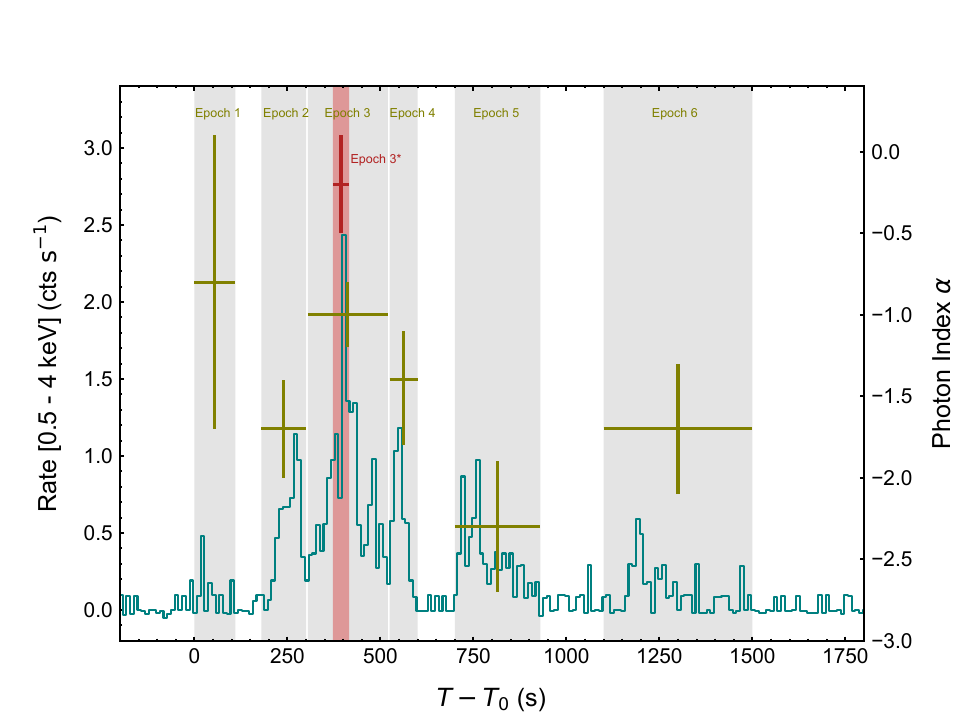}\put(0, 65){\bf b}\end{overpic} \\
\end{tabular}
\caption{\noindent\textbf{Soft X-ray image and light curve of the prompt emission by EP-WXT.} \textbf{a}, The image of the EP240315a in one of the 48 WXT CMOS chips on board the EP mission. The field of view of one CMOS is $9.3 \times 9.3$ square degrees. Two other bright X-ray sources are simultaneously detected. \textbf{b}, The light curve of the net count rate and spectral evolution of EP240315a in 0.5--4 keV. The grey shaded areas indicate the six epochs used for spectral analysis. The red box indicates the time window in which gamma-ray emissions were detected by Swift-BAT and Konus-Wind.}
\label{fig:wxt_img_lc}
\end{figure}

EP240315a was detected by the Wide-field X-ray Telescope (WXT) on board the Einstein Probe\cite{Yuan2022} (EP) in the soft X-ray band (0.5–4 keV) at $T_0 = $ 20:10:44 (UTC) on 15 March 2024\cite{EPWXTGCN} (Fig. \ref{fig:wxt_img_lc}a). An optical counterpart was discovered by the Asteroid Terrestrial-Impact Last Alert System (ATLAS) $\sim$1.1 h after the X-ray detection\cite{ATLASarXiv}. Subsequently, spectroscopic observations by the Very Large Telescope derived a redshift for this transient of $z=4.859$\cite{VLT2024GCN} and confirmed the cosmological origin of this source. The temporally and spatially coincident gamma-ray 
burst event GRB 240315C was discovered in the off-line analysis of the Burst Alert Telescope (BAT) on board the Neil Gehrels Swift Observatory\cite{DeLaunay2024GCN}, and Konus-Wind\cite{KWGCN35972}. At the redshift of $4.859$, the peak isotropic luminosity in the energy range of 0.5--4 keV reaches $(1.2 \pm 0.2) \times 10^{51}$ $\rm erg\,s^{-1}$, making EP240315a a unique gamma-ray burst triggered by a soft X-ray detector at a redshift of $z>4.5$.

The WXT light curve exhibits a multi-peaked structure (Fig. \ref{fig:wxt_img_lc}b). The $T_{90, \rm X}$ in the 0.5--4 keV band is 1,034$\pm$81 seconds, which is significantly longer than the $T_{90,\gamma}\sim 38 - 41$ seconds observed in the gamma-ray band (Table \ref{tab:obs_prob}). Remarkably, the soft X-rays triggered the WXT earlier than the gamma-rays recorded in the BAT and Konus-Wind by 372 seconds. This preceding time is significantly longer compared with the GRBs detected by BeppoSAX and High Energy Transient Explorer 2, of which the soft X-rays only precede the gamma rays by tens of seconds\cite{intZand1999,Frontera2000,Piro2005,Dalessio2006,
Galli2006,Vetere2007A&A,Sakamoto2005}. 
The whole spectrum of the prompt emission detected by WXT, integrated from $T_0$ to $T_0+$1,500 s, was fitted by an absorbed power law model with the photon index $\alpha = -1.4 ^{+0.1}_{-0.1}$ (the Galactic and intrinsic absorption column densities are fixed at $N_{\rm H}=4.4 \times 10^{20}$ $\rm cm^{-2}$ and $N_{\rm H, z}=5.9^{+4.9}_{-4.6} \times 10^{22}$ $\rm cm^{-2}$, respectively; Methods). The total unabsorbed fluence in 0.5--4 keV is $ (1.0 \pm 0.1) \times10^{-6}$ $\rm erg\,cm^{-2}$. The WXT count rate light curve can be divided into six epochs of pulses. The spectra fitted in each epoch exhibit a signature of evolution with a generally harder spectrum observed during the epoch of higher intensity (Fig. \ref{fig:wxt_img_lc}b and Table \ref{tab:wxt_spec}).   

\begin{figure}
\begin{minipage}[b]{0.5\textwidth}
    \begin{overpic}[scale=0.51]{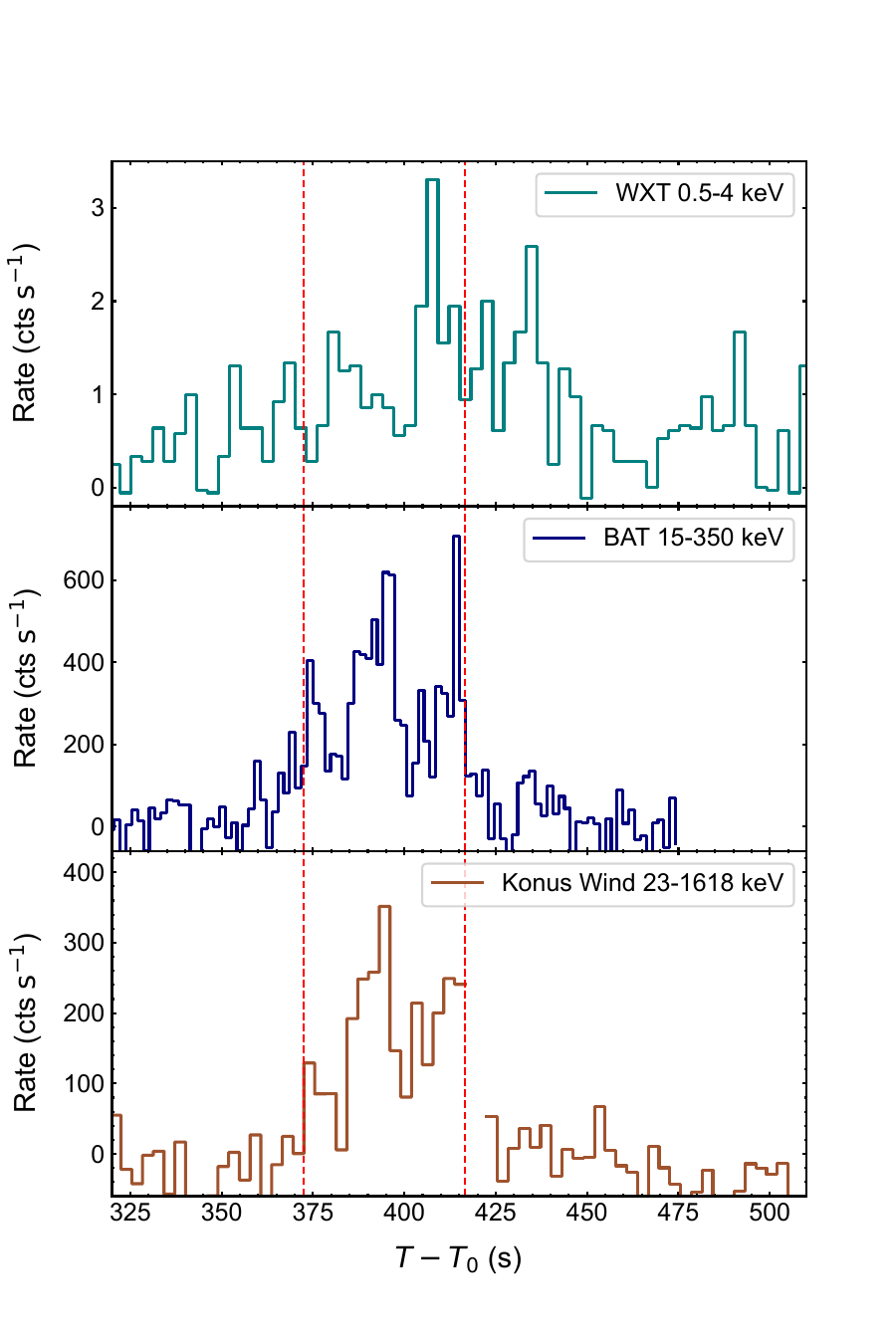}\put(1, 88){\bf a}\end{overpic}
\end{minipage}
\begin{minipage}[b]{0.6\textwidth}
    \begin{overpic}[scale=0.45]{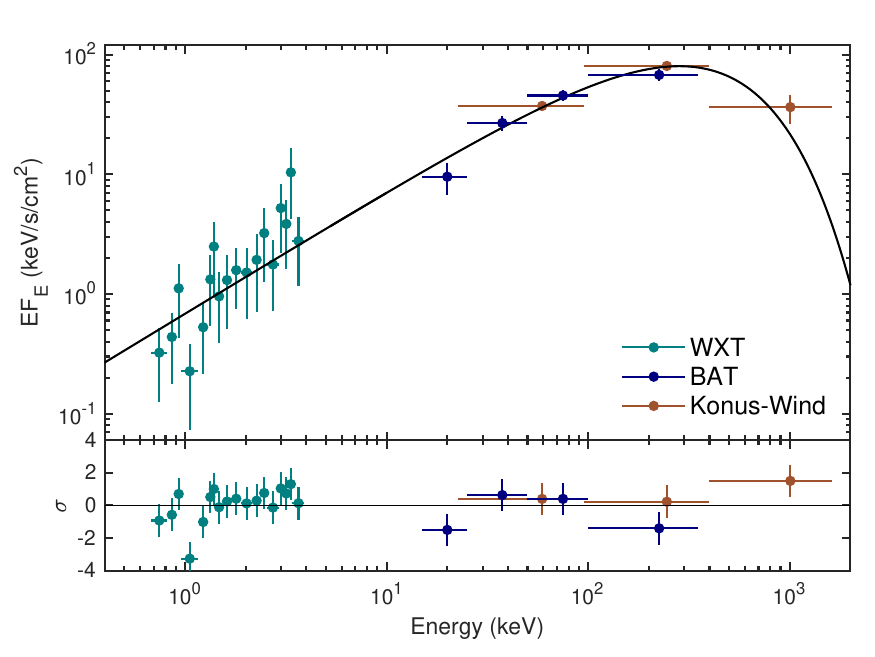}\put(0, 70){\bf b}\end{overpic} \\
    \begin{overpic}[scale=0.65]{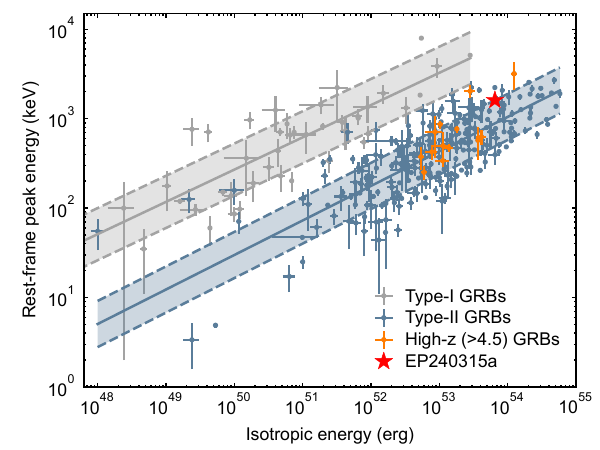}\put(0, 72){\bf c}\end{overpic} \\
\end{minipage}
\caption{\noindent\textbf{Temporal and Spectral behaviours of EP240315a/GRB 240315C during the joint detection by EP-WXT, Swift-BAT and Konus-Wind in Epoch 3*.} \textbf{a}, The multi-wavelength light curves of EP240315a/GRB 240315C by WXT, BAT and Konus-Wind. The time bins of WXT, BAT and Konus-Wind are 3 s, 1.6 s and 2.944 s, respectively. The X-ray and gamma-ray light curves exhibit similar temporal profiles, overlapping at their peaks, with the soft X-ray light curve generally broader. The red dashed lines denote the time interval of Epoch 3*. \textbf{b}, The joint spectral fitting of WXT, BAT and Konus-Wind spectra in Epoch 3* with an absorbed cutoff power law. The data points are shown with $1\sigma$ uncertainty and the black line represents the best-fit CPL model. \textbf{c}, The rest-frame peak energy versus isotropic energy correlation diagram. The burst locates at the high energy end with other high-$z$ GRBs, consistent with being a typical long GRB. The best-fit correlations and 1$\sigma$ scattering regions are presented by solid lines and shaded areas, respectively.}
\label{fig:joint_fit}
\end{figure}

\begin{figure}
\centering
\begin{tabular}{cc}
\begin{overpic}[width=0.45\textwidth]{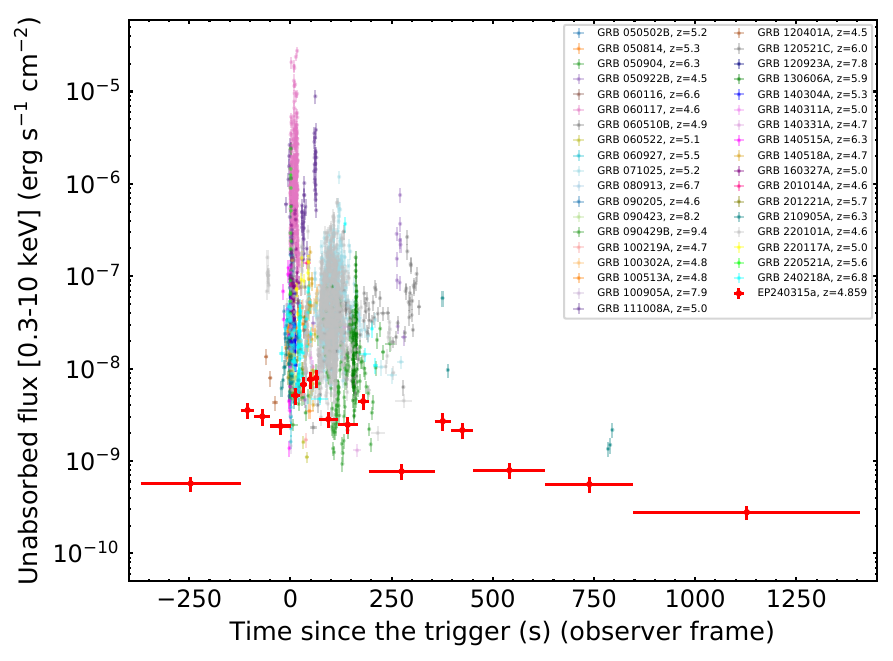}\put(0, 75){\bf a}\end{overpic} &
\begin{overpic}[width=0.43\textwidth]{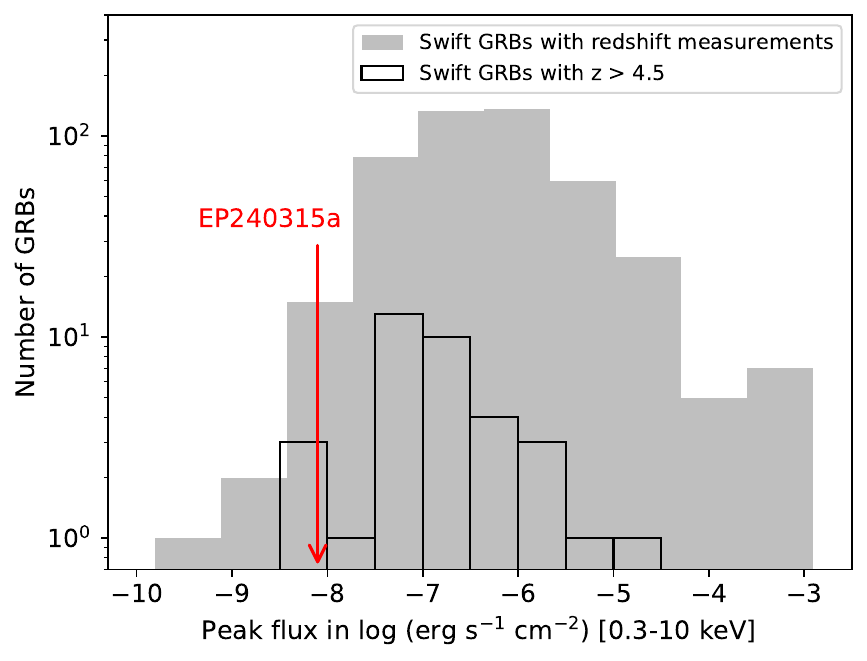}\put(0, 75){\bf b}\end{overpic} \\
\begin{overpic}[width=0.45\textwidth]{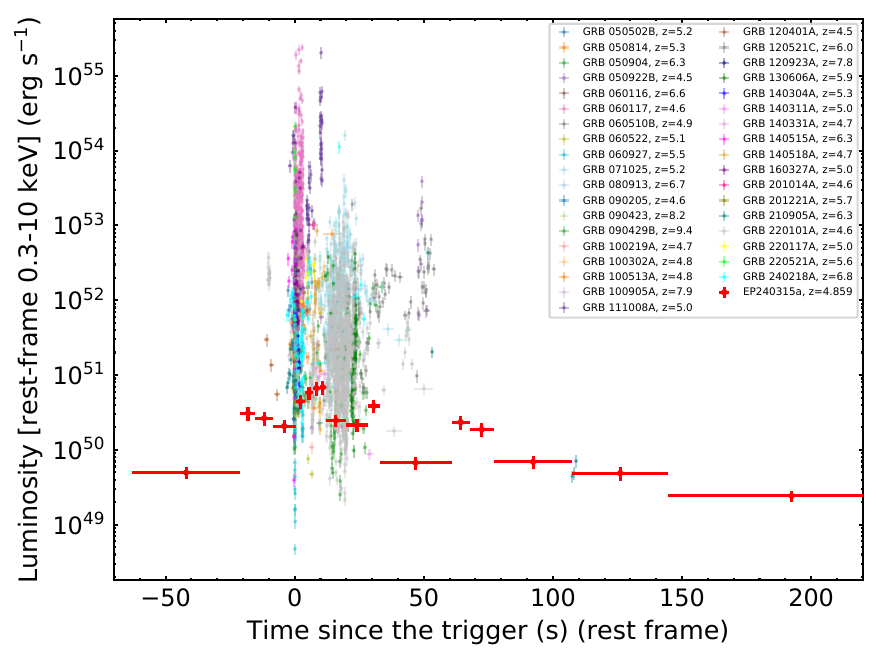}\put(0, 75){\bf c}\end{overpic} &
\begin{overpic}[width=0.45\textwidth]{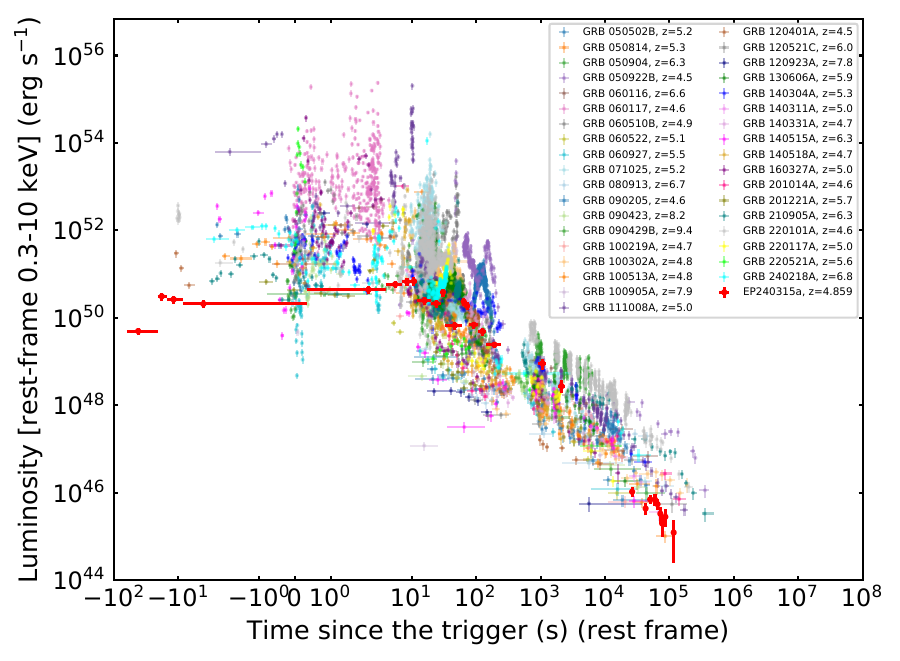}\put(0, 75){\bf d}\end{overpic} \\
\end{tabular}
\caption{\noindent\textbf{Comparison of EP240315a and Swift GRBs with $z>4.5$ in the soft X-ray band.} \textbf{a}, The unabsorbed flux light curve in 0.3--10 keV of EP240315a and Swift-BAT GRBs with $z > 4.5$ in the observer frame. The observed flux light curves of Swift-BAT GRBs are adopted from the Burst Analyser with a signal-to-noise ratio of at least 5\cite{Evans2010A&A}. The WXT prompt emission data are rebinned with a signal-to-noise ratio of 5. Some GRBs with the flux of $\sim10^{-9}$ $\rm erg\,cm^{-2}\,s^{-1}$ are detected by BAT in its optimal condition (e.g. GRB 060522 and GRB 140515A), which is higher than the sensitivity of WXT by one order of magnitude if $\alpha\sim-1$ (Methods).  \textbf{b}, Distribution of the peak flux in 0.3--10 keV of EP 240315a (red arrow) and Swift-BAT redshift-known GRBs sample (shaded histogram) and Swift-BAT GRBs subsample with redshift $z > 4.5$ (framed histogram). EP240315a is located at the fainter end of the distribution of the high-redshift sample. \textbf{c}, The k-corrected luminosity light curve in 0.3--10 keV of EP240315a and Swift-BAT GRBs with $z > 4.5$ in the rest frame.  \textbf{d}, Same as \textbf{c}, but including the late WXT and FXT observations of EP240315a and both the BAT and XRT light curves for the high-z GRBs.}
\label{fig:compare_SwiftGRBs}
\end{figure}

The gamma-ray detections coincide with the soft X-rays around the time of the peak, specifically between $T_0+372$ s and $T_0+416$ s (Epoch 3*). 
The X-ray and gamma-ray light curves exhibit similar temporal profiles, overlapping at their peaks, with the soft X-ray light curve generally broader (Fig. \ref{fig:joint_fit}a). 
A joint spectral fitting was performed using the spectra of WXT, BAT and Konus-Wind in Epoch 3*. The broadband spectrum within 0.5--1,618 keV can be well described by a cutoff power law (CPL) model with the photon index of $-0.97_{-0.05}^{+0.06}$ and peak energy of $283_{-47}^{+65}$ $\rm keV$ (Fig. \ref{fig:joint_fit}b). No significant additional component is required in the soft X-ray band. The equivalent isotropic energy in the 1--10,000 keV band is $6.4_{-0.8}^{+0.4}\times10^{53} $ $\rm erg$, placing this burst on the track of the Amati relation\cite{Amati2002A&A} towards the high energy end with other high-z GRBs (Fig. \ref{fig:joint_fit}c), consistent with being a typical long GRB. This indicates that the X-ray detected by EP-WXT is likely to have a common origin with the gamma-rays, being directly associated with the internal dissipation within the ejecta from the central engine. Notably, the observation window of the gamma-ray emission aligns with the peak flux and the hardest spectrum in X-rays, implying that the observed gamma-ray emission may be just a fraction of the total, with gamma-rays remaining undetectable during other observation epochs of WXT. The predicted fluxes in the gamma-ray band are consistent with the upper limits of BAT and Konus-Wind in the epochs without gamma-ray detection (Extended Data Fig. \ref{fig:fluxlimit}; Methods). WXT can detect even fainter pulses down to the flux level of  $\sim10^{-10}$ $\rm erg\,cm^{-2}\,s^{-1}$ with the durations of $\sim$100--200 s, though there is considerable probability for BAT to detect some epochs of EP240315a under its optimal conditions (Fig. \ref{fig:compare_SwiftGRBs}a; Methods).

\begin{figure}
\centering
\includegraphics[width=10cm]{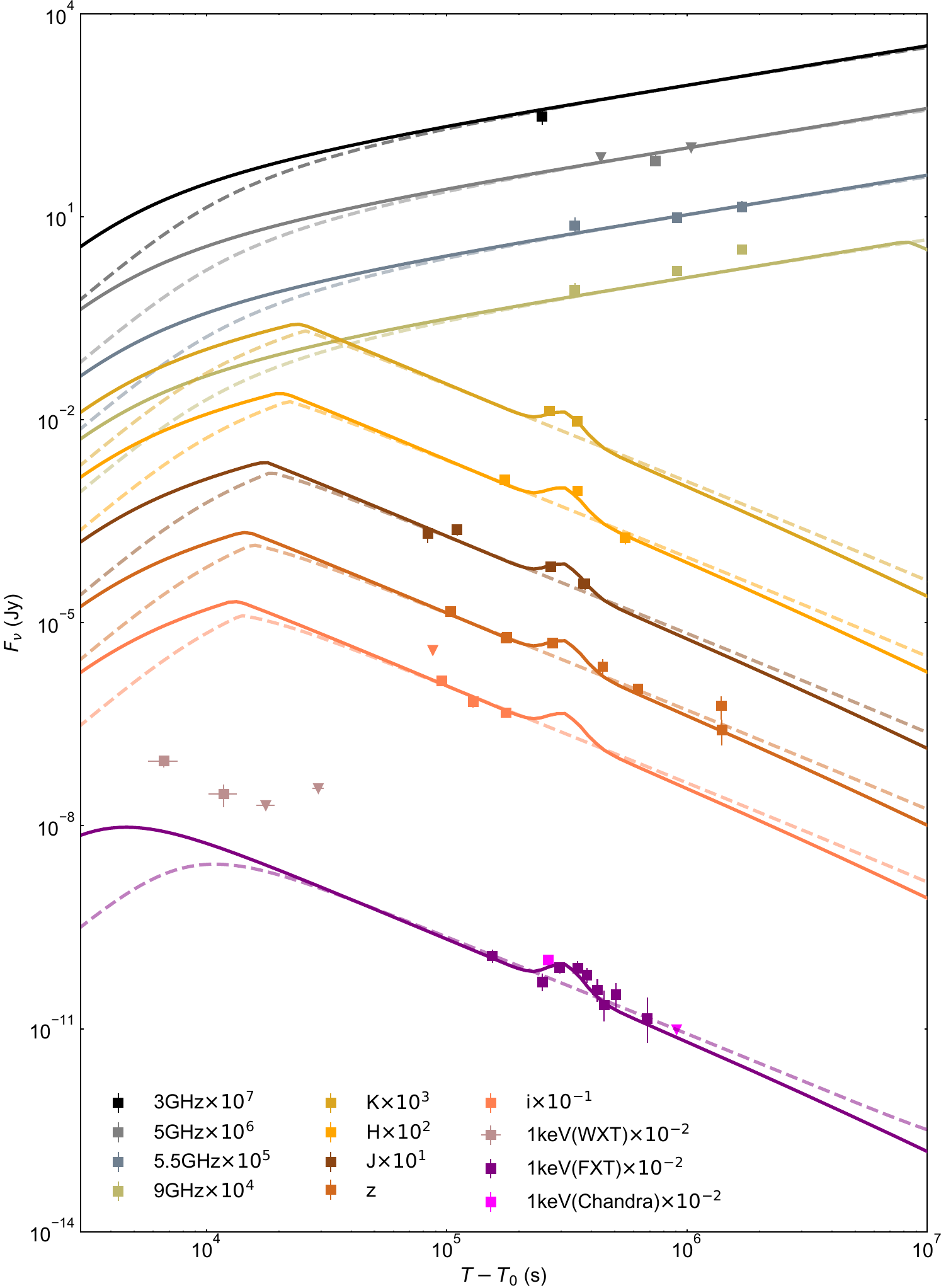}	
\caption{\noindent\textbf{The multi-wavelength afterglow light curves of EP240315a/GRB 240315C.} To distinguish data from different wavelengths, we use varying colors: squares indicate observational data points, while triangles represent upper limits. Utilizing a standard GRB afterglow model, we fitted the multi-wavelength data. Considering the strong Lyman alpha absorption due to high redshift, we limited our fitting to optical/infrared data with wavelengths not less than the \textit{i}-band. Presented here are the fitted data and results. Dashed lines represent the best-fit outcomes with the standard GRB afterglow model, whereas solid lines incorporate cases of late X-ray and optical re-brightening.}
\label{fig:aftgl_light_curve}
\end{figure}

Besides the long and bright emission in $T_{90,\rm X}$, WXT also detected faint signals in subsequent periods (from $T_0+5.7$ ks to $T_0+7.6$ ks and from $T_0+10.2$ ks to $T_0+13.4$ ks; Methods). The average flux from the late WXT observations indicates a simple power-law decline over time, with a slope of approximately $-2$. The photon index of the absorbed power law spectrum is $-1.9^{+0.8}_{-0.8}$ during these periods (Methods). The late X-ray observations by the Follow-up X-ray telescope (FXT)\cite{Chen2020SPIE}, also on board EP, started $\sim$42 hours after the WXT detection\cite{Chen2024GCN}. A faint X-ray signal was detected by the first nine out of ten follow-up observations until approximately 8 days after the burst (Methods). The Chandra X-ray observatory observed EP240315a twice, around $T_{0}+72$ hours and $T_{0}+10.4$ days, respectively\cite{Levan2024GCN.35963, Levan2024GCN.35982}. The first observation detected an X-ray source with a flux comparable to that detected by FXT around the same epoch. The second observation did not detect any source with a signal-to-noise ratio above 5, giving an upper limit of $4.1 \times 10^{-15}$ $\rm erg\,cm^{-2}\,s^{-1}$ in the 0.5--4 keV band. Combining the observations from FXT and Chandra, we find that the late-stage X-ray emission exhibits an overall trend of a power-law decay, with a distinct re-brightening structure observable between $T_0+200$ ks and $T_0+500$ ks (Fig. \ref{fig:aftgl_light_curve}). 

Following the prompt trigger by EP-WXT and the precise position localized by ATLAS, we launched a campaign to monitor the afterglow of EP240315a in the \textit{g, r, i, z, R, White, J, H, K} bands and in radio at 5 GHz, 5.5 GHz, and 9.0 GHz (Methods). We find that the standard GRB afterglow model\cite{meszaros1997} could well interpret the multi-wavelength follow-up data (Fig. \ref{fig:aftgl_light_curve}). 
The emissions detected by EP-WXT in the late observation after the prompt emission significantly deviate from the prediction of the afterglow model (Fig. \ref{fig:aftgl_light_curve}), suggesting that they more likely originate from the late activities of the central engine.
A possible re-brightening signature is also present in the optical data. The goodness-of-fit of the afterglow modelling is significantly improved when the late re-brightening signature in both X-ray and optical bands are incorporated (Methods).
The late-time X-ray re-brightening feature has been observed in some GRB afterglows\cite{Gao2017}, which is generally attributed to re-activation of the central engine or to off-axis observations of complex jet structures\cite{ZhangBB2014ApJ}.

Observations of the X-Ray Telescope (XRT) onboard Swift suggest that many GRBs have extended central engine activity, manifested through X-ray flares and shallow-decay plateaus following the main burst released in gamma-rays, typically lasting for hundreds of seconds and up to ten thousand seconds\cite{Burrows2005Sci, Zhang2006, Chincarini2007ApJ, Troja2007}. Nevertheless, rare simultaneous prompt emissions in the soft X-ray band were captured by XRT, only when Swift completes a rapid slew before the end of the prompt emission, or there is a bright precursor trigger in gamma-rays before the main pulses\cite{Romano2006,Oganesyan2017}. The X-ray characteristics of the entire prompt stage thus remain highly uncertain. This time, the EP-WXT observations of EP240315a clearly show that the central engine has already begun intense activity long before the detection of gamma-rays, with a whole prompt X-ray duration extending to 10--100 times longer than that of the gamma-rays. This casts doubt on the conventional practice of using the duration determined by gamma-ray observations to reflect underlying central engine activity. The early X-ray emissions may be caused by the prior shock breaking out from the star envelope\cite{Bromberg2012}, or some leading weak jets among the intermittent jets\cite{Lopez16,Geng16} from the unstable accretion of the central engine. Hence, our observations pose a challenge to previous analyses of the jet propagation and acceleration timescale based on the gamma-ray information only. 
Also, the discrepancy of the event zero point defined by these two bands would significantly influence the temporal behavior of the early emissions and relevant physical explanations, e.g., the decaying slopes of the early X-ray flares\cite{Liang2006,Jia17,Geng17}.
Therefore, we suggest that EP has ushered in an era of in-depth study of GRB central engine activity with the complementary observations of the gamma-ray detectors and depict the temporal features of early emissions more properly.

The rate of GRBs detected by Swift which end up to have their redshift measured above $z>4.5$ is approximately one every 8 months. EP240315a was discovered during the commissioning phase, approximately two months after the launch of EP. A comparison of EP240315a with the prompt emissions of high-redshift GRBs by Swift-BAT revealed that EP240315a exhibits a relatively faint peak flux in the soft X-ray band compared to other high-redshift sources (Fig. \ref{fig:compare_SwiftGRBs}a,b). EP240315a is also shown to exhibit a much longer timescale than the high-z GRBs observed by BAT through a continuous observation (Fig. \ref{fig:compare_SwiftGRBs}c). 
This has demonstrated the advantages of WXT in the detection of fainter transients with a large field of view and a sensitivity better than previous all-sky monitors by more than one order of magnitude. With a longer exposure, WXT can also detect the late X-ray observations which are related to the central engine activity (Fig. \ref{fig:compare_SwiftGRBs}d). Furthermore, the precise localization provided by WXT and the capability for early warning can significantly improve the efficiency of multi-band follow-up observations. This not only facilitates the identification of the early central engine activities and the multi-band characteristics of the afterglow, but also considerably enhances the probability of redshift measurements. The observation of EP240315a serves as a striking case of studying the soft X-ray properties of GRB prompt emission at the redshift above 4.5. Further investigation has shown that an EP240315a-like event can be well detected with EP-WXT at a redshift of 7.5 with a signal-to-noise ratio above 7 (Methods). This suggests that in the future, more GRBs with higher luminosity and redshifts will be discovered by EP. The era of soft X-ray detection of high-redshift gamma-ray bursts has officially begun with EP.

\clearpage

\clearpage

\clearpage

\clearpage

\clearpage

\begin{table*}
\centering
\scriptsize
\begin{threeparttable}
\caption{\textbf{Observational properties of the prompt emission.} All errors represent the 1$\sigma$ uncertainties.}
\label{tab:obs_prob}
\begin{tabular}{lccc}
\toprule
Observed Properties & EP-WXT  &  Swift-BAT &  Konus-Wind  \\
& 0.5--4 keV & 15--350 keV &  23--1,618 keV  \\  
\hline
Duration $T_{90}$ (s) & $1,034\pm81$ & $41.6\pm1.6$ &  $38\pm3$ \\
Photon Index $\alpha$\tnote{1} & $-1.4^{+0.1}_{-0.1}$  &  $-0.7^{+0.5}_{-0.4} $  & $-1.2^{+0.2}_{-0.1} $ \\
Peak Flux ($\rm erg\,cm^{-2}\,s^{-1}$) & $4.6^{+0.8}_{-0.7}\times10^{-9} $  &
$(4.9 \pm 0.5)\times10^{-7} $ &  
$8.7^{+1.7}_{-1.6}\times10^{-7} $\\
Peak Luminosity ($\rm erg\,s^{-1}$) & $(1.2\pm 0.2)\times10^{51}  $ &  $(1.2\pm 0.1) \times10^{53}  $ &  $(2.2 \pm 0.4)\times10^{53}$ \\
Total Fluence ($\rm erg\,cm^{-2}$)\tnote{1} & $(1.0\pm 0.1)\times10^{-6} $ &  $(6.6\pm 0.4)\times10^{-6} $ & $(1.8 \pm 0.2)\times10^{-5} $ \\
\midrule
&  \multicolumn{3}{c}{1--10,000 keV}\\
\hline
Isotropic Energy ($\rm erg $)\tnote{2} & \multicolumn{3}{c}{$6.4_{-0.8}^{+0.4}\times10^{53} $}\\
\bottomrule
\end{tabular}
\begin{tablenotes}
\footnotesize
\item[1] The photon indices and the total fluences are derived over the time intervals of [$T_0$, $T_0+1500$ s] for EP-WXT, and [$T_0+372$ s, $T_0+416$ s] for Swift-BAT and Konus-Wind. 
\item[2] The isotropic energy is derived from the joint fit of EP-WXT, Swift-BAT and Konus-Wind over the time interval of [$T_0+372$ s, $T_0+416$ s].
\end{tablenotes}
\end{threeparttable}
\end{table*}

\clearpage

\begin{table*}
\centering
\scriptsize
\begin{threeparttable}
\caption{\textbf{The fitting results and corresponding fitting statistics for the time-sliced prompt emission spectra.} All errors represent the 1$\sigma$ uncertainties.}
\label{tab:wxt_spec}
\begin{tabular}{cccccccc}
\toprule
Instruments & Spectrum & Time Interval & Model & Photon Index & $E_{\rm peak}$ & Flux \tnote{1} & CSTAT/(d.o.f) \\
& & (second) &  & $\alpha$ & (keV) &  ($\rm erg\,cm^{-2}\,s^{-1}$)\\
\hline
WXT & Total & 0--1,500 & PL & $-1.4^{+0.1}_{-0.1} $ & - &$6.1^{+0.4}_{-0.4}\times10^{-10} $ & 45.21/38 \\
& Epoch 1& 0--110 & PL & $-0.8^{+0.9}_{-0.9}$   &  - & $2.2^{+1.4}_{-1.0}\times10^{-10} $ & 11.57/1  \\
& Epoch 2& 180--305 & PL &$-1.7^{+0.3}_{-0.3}$ &  - &  $9.5^{+1.9}_{-1.6}\times10^{-10} $ & 21.29/17 \\
& Epoch 3& 305--525 & PL &$-1.0^{+0.2}_{-0.2}$ &  - &  $2.2^{+0.3}_{-0.3}\times10^{-9} $ & 42.09/46  \\
& Epoch 4& 525--600 & PL &$-1.4^{+0.3}_{-0.4}$  &  - &$1.3^{+0.3}_{-0.3}\times10^{-9} $ & 14.99/15 \\
& Epoch 5& 700--930 & PL &$-2.3^{+0.4}_{-0.4}$  &  - & $4.7^{+0.9}_{-0.7}\times10^{-10} $ & 23.83/22 \\
& Epoch 6& 1,100--1,500 & PL &$-1.7^{+0.4}_{-0.4}$  &  - & $2.3^{+0.5}_{-0.5}\times10^{-10} $  & 7.90/13\\
& Epoch 3*& 372--416 & PL &$-0.2^{+0.3}_{-0.3} $ &   - &$4.6^{+0.8}_{-0.7}\times10^{-9} $ & 11.38/16 \\
\midrule
BAT+Konus & Epoch 3* & 372--416 & CPL  & $ -1.1_{-0.1}^{+0.2}$   & $ 366_{-98}^{+150}$    & $3.9^{+0.8}_{-0.6}\times10^{-7}$   &  5.5/3 \\
WXT+BAT+Konus & Epoch 3* & 372--416 & CPL  & $-0.97^{+0.06}_{-0.05}$ &$283^{+65}_{-47}$ &$3.4^{+0.2}_{-0.4}\times10^{-7}$  & 24.32/22 \\
\bottomrule
\end{tabular}
\begin{tablenotes}
\footnotesize
\item[1] The averaged absorbed flux is derived in 0.5--4 keV for the WXT spectra, in 10--10,000 keV for the joint fitting of BAT+Konus spectra and in 1--10,000 keV for the joint fitting of WXT+BAT+Konus spectra.
\end{tablenotes}
\end{threeparttable}
\end{table*}

\clearpage

\section*{Methods}
\subsection{Observations and data reductions.\\} 
\noindent\textbf{EP-WXT.} The Einstein Probe (EP)\cite{Yuan2022} launched on 9 Jan 2024, is a mission of the Chinese Academy of Sciences (CAS), in collaboration with the European Space Agency (ESA) and the Max Planck Institute for extraterrestrial Physics (MPE), Germany, dedicated to time-domain high-energy astrophysics. Wide-field X-ray Telescope (WXT) is one of the two main payloads of EP. It employs novel lobster-eye micro-pore optics (MPO) to enable a large instantaneous field of view (FoV) of 3,600 square degrees and a sensitivity of $\sim2.6\times10^{-11}$ $\rm erg\,cm^{-2}\,s^{-1}$ in 0.5--4 keV with 1\,ks exposure. 
EP240315a was detected by the WXT during the commissioning phase of EP and a long calibration observation of FXT of Vela SNR from UTC 2024-03-15T18:00:22 to 2024-03-16T05:08:19 (ObsID 13600005120). Due to Earth occultations and SAA passages, this observation was composed of nine segments. The onset and prompt emission of EP240315a was covered by the second segment ($T_0$ is defined at 2024-03-15T20:10:44), and the following four segments also covered this source. The source in two segments, from $T_0+5.7$ ks to $T_0+7.6$ ks and from $T_0+10.2$ ks to $T_0+13.4$ ks, was detected with a signal-to-noise ratio higher than 3$\sigma$. Other segments did not detect any signal above the 3$\sigma$ level and were therefore only able to constrain the upper limit. This source also triggered the on-board processing unit and the ground center received the real-time alert after 10 s (the autonomous follow-up was not enabled at that time).

The X-ray events were selected and calibrated using the data reduction software and calibration database (CALDB) designed for WXT (Liu et al. in prep.). The CALDB is generated based on the results of the on-ground calibration experiments (Cheng et al. in prep), the procedure of which has been applied to a prototype of WXT instrument\cite{Cheng2024}. The position of each photon was converted to celestial coordinates (J2000). The energy value of each event is calculated according to the bias and gain stored in CALDB. After bad/flaring pixels were flagged, single-, double-, triple-, and quadruple-events without anomalous flags were selected to form the cleaned event file. The image in the 0.5--4 keV range was extracted from the cleaned events (Fig. \ref{fig:wxt_img_lc}a) . The light curve and the spectrum of the source and background in a given time interval were extracted from a circle with a radius of 9 arcmin and an annulus with the radii of 18 arcmin and 36 arcmin, respectively. Since the peak count rate is only 0.125 ct/frame in the source region, the pile-up effect is negligible in the WXT data.

\noindent\textbf{Swift-BAT.} BAT is a coded mask imager sensitive to hard X-rays and gamma-rays in the 15 keV to 350 keV energy range. During the peak emission time of EP240315a, the Swift satellite was slewing and BAT's GRB triggering system was also turned off, which resulted in no onboard detection of EP240315a. During slew, BAT's survey data products are also not produced; therefore sky images cannot be produced during this period. BAT rates data is available during this period which are the total count rates across all or large portions of detectors binned across four energy ranges. The data product analysed in this paper is the ``quad-rates" data, which bins the count rates across the four quadrants of the detector plane at a timing resolution of 1.6 s. The quad-rates data is also binned across four energy bins: 15--25~keV, 25--50~keV, 50--100~keV, and 100--350~keV. Had there been a prompt alert for EP240315A, time-tagged event data could have been available via the Gamma-Ray Urgent Archiver for Novel Opportunities (GUANO) \cite{GUANO} system that allows BAT to save and deliver time-tagged event data to the ground around a time of interest. 

The peak emission of EP240315a is clearly seen in the BAT quad-rates data, starting at $\sim T_0$ + 370 s. The count rate light curve is difficult to interpret as the Crab is travelling through BAT's coded field of view during the slew. Using the detector responses developed for the Non-Imaging Transient Reconstruction and Temporal Search (NITRATES)\cite{NITRATES}, the rate expectation from the Crab was calculated for each detector quadrant, energy bin and time bin around the peak emission period and subtracted from the light curve. For the full background model, a linear fit as a function of time was also performed on the count rates after the Crab contribution was subtracted for each quadrant and energy bin using the times before and after the emission was detected. With the full background model (linear fit plus Crab) subtracted from the quad rates data, a $T_{90,\gamma}$ of 41.6 s is obtained over the 15--350 keV energy range. 

Using the NITRATES likelihood framework, a similar analysis was developed to be used on quad-rates data, where instead of having likelihood and model expectation for each detector, the model expectation is summed over each quadrant of detectors and the log-likelihood (LLH) for each quadrant and energy bin is summed to find the total LLH. Using the linear plus Crab background model, a transient point source is searched for by maximising the LLH over the signal parameters: sky position, flux normalisation and spectral shape. After scanning the likelihood over a grid of positions over the sky and profiling over several spectral templates, Bayes theorem is used to calculate the posterior probability density as a function of position for the transient point source. We find a 50\% statistical credible region with a solid angle of 1,060 square degrees (90\% credible area of 6,400 square degrees), with the position of EP240315a firmly inside. 

\noindent\textbf{Konus-Wind.} The Konus-Wind instrument\cite{Aptekar_1995SSRv_71_265} (KW) is a gamma-ray spectrometer consisting of two identical detectors, S1 and S2, which observe the southern and northern ecliptic hemispheres, respectively. KW was continuously observing the whole sky during the entire duration of EP240315a. About 370~s after $T_0$, the long-duration GRB~240315C was detected by KW in the waiting mode\cite{KWGCN35972}. In this mode, count rates with a coarse time resolution of 2.944 s are recorded in three energy bands: G1 (23--96~keV), G2 (96--396~keV), and G3 (396--1,618~keV). The burst was observed over the high and variable background due to a solar particle event. A Bayesian block analysis of the KW waiting mode data in the 20--400~keV band reveals a $\sim 21 \sigma$ count rate increase in the interval from $T_0+372$~s to $T_0+417$~s. The burst light curve shows a multi-peaked structure with the brightest peak around $T_0+392$~s (we note a 6 s gap in the data starting from $T_0+417$ s).

Exploiting the difference in the arrival time of the gamma-ray signals at KW and Swift-BAT provided constraints on the burst localization, which was consistent with the EP240315a position\cite{Svinkin_2024GCN_35966_IPN}.

\noindent\textbf{EP-FXT.} The Follow-up X-ray Telescope\cite{Chen2020SPIE} (FXT) is one of the main payloads of EP in 0.3--10 keV. It consists of two modules (FXT-A and FXT-B), each containing 54 nested Wolter-I paraboloid-hyperboloid mirror shells. Each module is equipped with a PN-CCD as the focal
plane detector. The PN-CCD has an integrated image and a frame-store area, both with 384 × 384 pixels. The PN-CCD of FXT offers three readout modes, Full Frame mode (FF), Partial Window mode (PW) and Timing mode (TM). The combined effective area of two FXT modules, as measured in on-ground calibration, reaches $\sim$700~cm$^2$ at around 1~keV. The FXT is designed for quick follow-up observations and more precise positioning of transients, and for observations of targets of opportunity (ToO). 

All X-ray data from both the source and background regions were processed using the FXT data analysis software (fxtsoftware\;v1.05). The process involved particle event identification, PI conversion, Grade calculation and selection (grade $\leq12$), bad and hot pixels flag, selection of good time intervals using housekeeping file. The pipeline finally resulted in the creation of clean event files and energy spectrum files and response files.

The EP-FXT carried out ten follow-up observations of EP240315a, spanning from 42 hours to approximately ten days after the burst. The cumulative observation time exceeded 80,000 seconds, with details in the Extended Data Table \ref{tab:fxt_obsinfo} below. During these ten observations, both FXT-A and FXT-B observed EP240315a in full-frame mode. The first nine observations were conducted with a thin filter, while the tenth observation (ObsID: 08503014656) was conducted with a small aperture filter. As the FXT is still undergoing in-orbit calibration, the calibration of the effective area has not been finalised and may contain a systematic error of up to 10\%.

\noindent\textbf{Optical and Near-Infrared.} \textcolor{black}{We performed our multiband photometric follow-up with the following instruments: the Beijing Faint Object Spectrograph and Camera (BFOSC) on the Xinglong 2.16m Telescope\cite{Fan2016PASP}, the half meter telescope (HMT; 0.5 m located at Xingming Observatory, China), the Alhambra Faint Object Spectrograph and Camera (ALFOSC) on the Nordic Optical Telescope (NOT; 2.56 m at the Roque de los Muchachos observatory, La Palma, Spain), the Near Infrared Camera Spectrometer (NICS) on the Telescopio Nazionale Galileo (TNG; 3.58 m telescope located on the Island of San Miguel del La Palma in the Canary Islands), the Espectr\'{o}grafo Multiobjeto Infra-Rojo (EMIR)\cite{Garzon2022emir}, the High PERformance CAMera (HiPERCAM)\cite{Dhillon2021HiPERCAM} and the Optical System for Imaging and low-Intermediate-Resolution Integrated Spectroscopy (OSIRIS)\cite{Cepa2003OSIRIS} mounted on the Gran Telescopio CANARIAS (GTC; 10.4 m telescope). The X-shooter\cite{Vernet2011Xshooter} and the FOcal Reducer/low dispersion Spectrograph 2 (FORS2) mounted on the European Southern Observatory (ESO) Very Large Telescope (VLT; 8.4 m located at the Cerro Paranal, Chile). The LBT Near Infrared Spectroscopic Utility with Camera and Integral Field Unit for Extragalactic Research (LUCI, formerly LUCIFER) on the the Large Binocular Telescope (LBT; 8.4m located at Mount Graham, USA). After standard data reduction by known package (eg, IRAF\cite{Tody1986SPIE}) and astrometric correction\cite{Lang2010}, the optical apparent photometric were calibrated with the Pan-STARRS DR2 catalog\cite{Chambers2016arXiv,Flewelling2020ApJS} while the near-infrared data were calibrated with the 2MASS catalogue \cite{Skrutskie2006TwoMASS} or VISTA Kilo-degree Infrared Galaxy (VIKING) Survey\cite{Edge2013VISTA}. The Johnson-Cousin filter is calibrated with the converted magnitude from the Sloan system\footnote{\url{https://www.sdss.org/dr12/algorithms/sdssUBVRITransform/\#Lupton}}. The details of these observations and the photometric results are presented in Extended Data Table \ref{tab:phot}. Meanwhile, we collected the observations reported on the General Coordinates Network (GCN) to perform a joint analysis. The collected results are also listed in Extended Data Table \ref{tab:phot}.}

\noindent\textbf{Radio.}
Observations with the Australia Telescope Compact Array (ATCA) were carried out under program CX564 (PI: Troja) at the centre frequencies of 5.5 and 9 GHz with a bandwidth of 2 GHz. For these runs, the primary calibrator was 1934-638, and the phase calibrator 0941-080. Data were reduced using standard tasks within the MIRIAD package\cite{sault1995}. The target was clearly detected at all epochs, our measurements are reported in Extended Data Table \ref{tab:phot}. Observations were also performed with the enhanced Multi Element Remotely Linked Interferometer Network (e-MERLIN) under a joint Director’s Discretionary Time proposal (DD17003 PI Piro and DD17004 PI Rhodes) at a centre frequency of 5 GHz. Observations were taken at two different epochs, each epoch covering two consecutive nights. We used 3C286 for flux scale calibration and 0933-0819 for complex gain. The beam size was $275\times104$ and $150\times30$ milliarcsecond, and the RMS were 15 and 21 $\mu$Jy/beam in each epoch, respectively. The data were reduced with the e-MERLIN pipeline\cite{Moldon2021} and imaged with CASA. We did not detect the transient at either individual epoch. We, therefore, concatenated the two e-MERLIN epochs, achieving a reduced rms of 11 $\mu$Jy/beam with a beam size of $175\times32$ milliarcsecond, resulting in the detection of the transient. The 5-$\sigma$ upper limits of the individual epochs and the measurement in the combined dataset are reported in Extended Data Table \ref{tab:phot}.

\subsection{Spectral Analysis\\}
\noindent\textbf{EP-WXT and FXT.} The integrated spectrum of the WXT prompt emission and the spectra of the six epochs defined in Fig. \ref{fig:wxt_img_lc}b were fitted in \textit{XSPEC} by an absorbed power law model {\it tbabs*ztbabs} {\it *powerlaw}, where the first and second components are responsible for the Galactic absorption and the intrinsic absorption ($N_{\rm H}$), and the third one is a power law function, $N(E)=KE^{\alpha}$, in the observer's frame. The column density of the Galactic absorption in the direction of the burst is fixed at $4.4 \times 10^{20}~{\rm cm^{-2}}$ \cite{Willingale2013} and the redshift is fixed at 4.859. 
 Assuming no significant variation in the absorption column density within the prompt emission of EP-WXT, we performed a simultaneous fitting of the spectra of the six epochs. A time-averaged absorption of $N_{\rm H}=5.9_{-4.6}^{+4.9} \times 10^{22}~{\rm cm^{-2}}$ is obtained and thus fixed to determine the photon index of each epoch spectrum. The significant intrinsic absorption is consistent with other high-redshift GRBs\cite{Salvaterra2015JHEAp,Rossi2022A&A}. The fitted results and corresponding fitting statistics are shown in Table \ref{tab:wxt_spec}. 

The spectrum of the first segment of the WXT observation following the prompt phase was also fitted with the same absorbed power law model. The fitting gives an intrinsic absorption of $N_{\rm H,z}=2.5^{+13.7}_{-2.5}\times10^{22}~{\rm cm^{-2}}$ and a photon index of $-1.9^{+0.8}_{-0.8}$. A count-rate-to-flux conversion factor was derived from the fitting and applied to derive the flux of the second segment of the WXT observation. The flux upper-limits of the rest of the two segments are inferred with the same spectral model. 

FXT has performed ten observations of the afterglow. A persistent source 8.7 arcsec away from the GRB is revealed by the two Chandra observations, which cannot be resovled by FXT. The non-detection of the afterglow by Chandra's second observation suggested that the last FXT observation taken at a similar time only detected the persistent source. No obvious flux variation is seen in the 10th FXT observation, and the flux of this observation is consistent with the flux of the persistent source observed by Chandra. Therefore, the spectra of the first 9 FXT observations are subtracted by the spectra of the 10th observation to get the net spectrum of the afterglow. The net spectra can be fitted simultaneously by the same model applied to the WXT spectra with an intrinsic absorption of $N_{\rm H,z}=1.0_{-1.0}^{+1.3}\times 10^{23}~{\rm cm^{-2}}$ and a photon index of $-1.9^{+0.5}_{-0.5}$.

In order to derive the isotropic energy and the luminosity in the energy range [$E_1$, $E_2$] in the rest frame from the observed energy range [$e_1$, $e_2$], we perform a k-correction which can be expressed as
\begin{equation}
k=\frac{\int_{E_1/(1+z)}^{E_2/(1+z)} E N(E) dE}{\int_{e{1}}^{e_{2}} E N(E) dE}.
\label{kcorrection}
\end{equation}

\noindent\textbf{Swift-BAT.}
The same likelihood analysis discussed in the data reduction section was used to perform a spectral fit with the position fixed at EP240315a's localisation at the same period as the joint spectral fit, $T_0$ + 372 s to $T_0$ + 416 s. Fitting with a cutoff power-law resulted in a photon index of $-0.7^{+0.5}_{-0.4}$, a peak energy of $200^{+100}_{-50}$ keV, and an average 15--350~keV flux of $1.5^{+0.1}_{-0.1} \times 10^{-7}$ erg cm$^{-2}$ s$^{-1}$. The error bars are 68\% and found using Wilks' theorem\cite{Wilks}. For the joint fitting a time averaged detector response was constructed and summed over the four quadrants to analyze a single spectra in four energy bins. 

The peak flux was found by maximising the LLH over the flux with the peak energy and photon index fixed to 200 keV and -0.7 respectively for each 1.6 s time bin. The peak 1.6 s flux was found starting at $T_0$ + 412.6 s at a flux of $4.9^{+0.5}_{-0.5}\times10^{-7}$ erg cm$^{-2}$ s$^{-1}$. The energetics were found using the peak flux and the fluence during the joint fitting period along with the redshift, $z = 4.859$. We found the peak isotropic luminosity to be $1.2^{+0.1}_{-0.1}\times10^{53}$ erg s$^{-1}$ and the total isotropic energy to be $1.6^{+0.1}_{-0.1}\times10^{54}$ erg.

\noindent\textbf{Konus-Wind.}
The emission evolution was explored using three-channel spectra, constructed from the counts in the G1, G2 and G3 energy bands in the four intervals corresponding to the Bayesian blocks and for the total burst interval (Extended Data Table~\ref{tab:kw_spec}). Details on Konus-Wind three-channel spectral analysis can be found elsewhere\cite{Svinkin_2016ApJS_224_10,Tsvetkova_2021}.
We performed the spectral analysis in XSPEC, version 12.11.1, using
the following spectral models: a simple power law, and a custom exponential
cutoff power-law (CPL) parameterized by the peak energy of $\nu F(\nu)$ spectrum
and with the energy flux as the model normalization.

Since a CPL fit to a three-channel spectrum has zero degrees of freedom (and, in
the case of convergence, $\chi^2 = 0$), we do not report the statistic
for such fits. The 68\% confidence intervals of the parameters were
calculated using the command \texttt{steppar} in XSPEC.

The total energy fluence $S$ was derived using the 10~keV--10~MeV energy flux of the best-fitting CPL spectral model for the total interval. 
The peak flux $F_\mathrm{peak}$ was calculated on the 2.944-s scale using the energy
flux of the best-fitting CPL model to the three-channel spectrum for the Bayesian block containing
the peak count rate interval ($T_0+387.236$~s to $T_0+396.068$~s). The burst total energy fluence is $1.9^{+0.4}_{-0.3}\times 10^{-5}$~$\rm erg\,cm^{-2}$,
and the 2.944-s peak energy flux is $9.1^{+2.2}_{-1.9}\times 10^{-7}$~$\rm erg\,cm^{-2}\,s^{-1}$, both in the 10~keV--10~MeV energy range.

Using $z = 4.859$ and the values of $S$ and $F_\mathrm{peak}$, we estimate the rest-frame energetics of the burst in the Konus-Wind band. 
Assuming a flat $\Lambda$CDM cosmology with $H_0 = 67.7$~km~s$^{-1}$~Mpc$^{-1}$ $\Omega_M = 0.315$,
the total isotropic energy $E_\mathrm{iso}$ is $8.2_{-1.2}^{+1.6}\times10^{53}$~erg and 
the peak isotropic luminosity $L_\mathrm{iso}$ is $2.3_{-0.5}^{+0.5}\times10^{53}$~erg~s$^{-1}$.

\subsection{Joint Spectral Fitting.}
We performed joint spectral fitting in Epoch 3* using the spectra of WXT, BAT and Konus-Wind. The spectra can be well fitted by an absorbed cutoff power law {\it tbabs*ztbabs*cutoffpl}  (CPL):
\begin{equation}
N(E)=A\Big(\frac{E}{100\,{\rm keV}}\Big)^{\alpha}{\rm exp}\Big[-\frac{E(2+\alpha)}{E_{\rm peak}}\Big],
\end{equation}
where the peak energy $E_{\rm peak}$ is related to the cutoff energy $E_{\rm c}$ through $E_{\rm peak}=(2+\alpha)E_{\rm c}$.

The intrinsic absorption is fixed at the value determined by the integrated spectrum of WXT to avoid any bias introduced by the uncertainty of cross-calibration between the WXT spectrum and gamma-ray spectra. 

A constant factor of 0.8 is introduced for the spectra model of BAT to account for the systematic difference between Konus and BAT\cite{Sakamoto2011PASJ,Tsvetkova_2021}.
According to the observation of Crab during the in-flight calibration campaign, the systematic uncertainty of effective area of WXT is about 10 percent (Cheng et al. in prep). Therefore, we fixed the constant factor for WXT and Konus as unity to avoid missing any additional feature in the WXT band.

There is no significant evidence for any additional component in the WXT band, though the photon index obtained from the joint fitting is softer than that only using WXT spectrum, which indicates there could be a spectral break $\sim$10 keV\cite{Oganesyan2017}. However, the exact shape cannot be constrained by the current data due to the gap and the uncertainty of the cross-calibration between the WXT spectrum and gamma-ray spectra.   

\subsection{Upper limit and sensitivity estimate\\}
During epochs 1, 2, 4, 5 and 6 (as listed in Table \ref{tab:wxt_spec}) no emission was detected by BAT. We calculate upper limits during those epochs assuming the best fit spectral shape found in the joint fitting, a cutoff power-law with $E_{\rm peak} = 366$ keV and the corresponding best-fit $\alpha$ for EP-WXT from each epoch. When searching for emission in rates data over a varying background,  BAT is only sensitive to impulsive emission over short time scales; thus, upper limits are determined for a 16 s timescale. We set upper limits over the entire duration of each epoch by finding a series of 16 s upper limits and taking the highest limit found for each epoch (Extended Data Fig. \ref{fig:fluxlimit}). 

EP240315a was outside the coded field of view of BAT at epochs 1, 2, and 4-6. Had EP240315a happened to be in the centre of the field of view of BAT (where its peak sensitivity is) it would have had an increased sensitivity to the long durations of the epochs via its imaging capabilities. We find BAT's sensitivities at these epochs, given an $E_{\mathrm{peak}}$ of 366 keV and the best-fit $\alpha$ for EP by first taking a long exposure of BAT event data from earlier in the day and creating images with the same exposure as each epoch's duration. We then find the image noise value at the centre of the field of view. Finally, we create a detector response for this position and find the flux needed to create enough counts to have a signal-to-noise ratio of 5. The resulting sensitivities in 15--350 keV arrange from 2 to $8\times10^{-9}$~erg\,cm$^{-2}$\,s$^{-1}$. 

During epochs 1, 2, 4, 5, and 6 (as listed in Table \ref{tab:wxt_spec}) no emission was detected by Konus-Wind. For each epoch, we estimate an upper limit ($5\sigma$ CL)
on the 15--350~keV flux at 15~s timescale to be $1\times10^{-7}$~erg\,cm$^{-2}$\,s$^{-1}$, assuming cutoff power-law model with $E_{\mathrm{peak}} = 366$~keV and the corresponding best-fit $\alpha$ for EP-WXT from each epoch (Extended Data Fig. \ref{fig:fluxlimit}).

The extrapolated flux in 15--350 keV is calculated assuming the same photon index in each epoch and a high energy cutoff obtained by the joint fitting. The predicted fluxes in Epochs 1, 2, 4, 5 and 6 are compared with the upper limits of BAT and Konus-Wind in Extended Data Fig. \ref{fig:fluxlimit}.

Since EP240315a is detected near the edge of the FoV of WXT, the effective area at 1 keV is about $70\%$ of the value at the center of FoV. 
To compare with the sensitivity of BAT in its optimal condition, the $5\sigma$ sensitivity of WXT (converted to 15--350 keV) is estimated assuming the same photon index in each epoch, fixed $E_{\mathrm{peak}} = 366$~keV and the maximum effective area of WXT. The resulting sensitivities arrange from  $\sim$$10^{-10}$~erg\,cm$^{-2}$\,s$^{-1}$  
 ($\alpha\!\sim\!-2$) to $10^{-8}$~erg\,cm$^{-2}$\,s$^{-1}$ ($\alpha\!\sim\!-1$).

\subsection{Amati relation\\} We utilize the previously collected samples of 45 type I and 275 type II redshift-known GRBs to investigate the correlations (also known as Amati relation) between rest-frame peak energy, $E_{\rm peak,z}$, and isotropic energy, $E_{\rm\gamma,iso}$, for both type I and type II GRBs\cite{Amati2002A&A, Zhang2009ApJ, Minaev2020MNRAS}. Such correlations can be fitted by the linear models expressed as ${\rm log}E_{\rm peak,z}=b + k{\rm log}E_{\rm\gamma,iso}$. The likelihood function is determined using the orthogonal-distance-regression method\cite{Lelli2019MNRAS}, incorporating an additional uncertainty term, $\sigma_{\rm int}$, to account for the intrinsic scatter of data along the perpendicular direction. The posterior probability distributions of parameters generated by Python module \emph{emcee}\cite{Foreman-Mackey2013} give the best-fitting parameters with 1$\sigma$ uncertainties to be $k_{\rm I}=0.36_{-0.05}^{+0.04}$, $b_{\rm I}=-15.6_{-2.1}^{+2.5}$ and ${\rm log}\sigma_{\rm int,I}=-1.29_{-0.13}^{+0.13}$ for type I GRBs, and $k_{\rm II}=0.39_{-0.02}^{+0.02}$, $b_{\rm II}=-17.8_{-1.0}^{+0.9}$ and ${\rm log}\sigma_{\rm int,II}=-1.42_{-0.05}^{+0.05}$ for type II GRBs. The two types of GRB samples, as well as the corresponding correlations and 1$\sigma$ intrinsic scattering regions, are presented in Fig. \ref{fig:joint_fit}c.

\subsection{Multi-wavelength afterglows fitting\\} 

Here we applied the standard GRB afterglow model to fit the multi-band follow-up data, by performing the $\chi^2$ objective function minimization procedure with the Markov Chain Monte Carlo method through the \emph{emcee} code\cite{Foreman-Mackey2013}. The dynamic evolution of a GRB jet is calculated based on Ref.\cite{huang2000} and the radiation flux density is calculated based on Ref.\cite{sari1998}.

We fitted the multi-band data with two approaches: one accounting for late-time X-ray/optical brightening, the other not. The rebrightening signature is described by a smooth broken power-law function\cite{yi2016} 

\begin{equation}
F_{\nu}\left(t\right)=F_{0}\nu^{\beta}\left[\left(\frac{t}{t_{\rm b}}\right)^{\alpha_1\omega}+\left(\frac{t}{t_{\rm b}}\right)^{\alpha_2\omega}\right]^{-1/\omega}.
\end{equation}
where $\beta$ is the spectral index, $\alpha_1$ and $\alpha_2$ are the temporal slopes, $t_{\rm b}$ is the break time and the $\omega$ represents the sharpness of the peak. The best fit parameters of two fitting approaches are shown in Extended Data Table \ref{parameters}. In Fig. \ref{fig:aftgl_light_curve} and Extended Data Fig. \ref{fig:MCMC1}-\ref{fig:MCMC2}, we show our best-fitting results for the light curve and the corresponding corner plot of the posterior probability distribution for the fitting. Taking into account the strong Lyman alpha absorption caused by the high redshift, we only used optical/infrared data with wavelengths not less than the \textit{i}-band in the fitting process.

From the fitting results, it can be seen that the standard GRB afterglow model can well interpret the multi-band follow-up observation data of EP240315a (Fig. \ref{fig:aftgl_light_curve}). The emissions of the two periods around $T_0+10^5$ s by EP-WXT significantly deviate from the prediction of the afterglow model, suggesting that they are more likely originated from the late activities of the central engine. The Bayesian information criterion (BIC), defined as BIC$=\chi^2+k\text{ln}N$, is employed to compare and select models, where $k$ is the number of free parameters of the model, and $N$ is the number of data points. According to the BIC criterion, incorporating the late-time X-ray re-brightening significantly enhances the fitting effect. Due to the lack of optical/infrared observation data in the very early and late stages, some afterglow parameters, e.g., the half-opening angle of the GRB jet,  are not well constrained. Nevertheless, this does not undermine the conclusion that the standard afterglow model adequately explains multi-band follow-up data of EP240315a.

\subsection{Prospects for high-$z$ detection\\}
We investigate the potential of WXT in detecting high-$z$ GRB for EP240315a-like events. With the observed luminosity, X-ray spectrum, light curve and absorption, we place it at different redshifts to evaluate the detection capability with EP-WXT. 

For EP240315a-like events, we simulate a hypothesized source and a background spectrum using \textit{fakeit} in \textit{XSPEC} by assuming the best-fitting continuum model of EP240315a. The same exposure time, redistribution matrix and auxiliary file for each of the sources at different redshifts are used in these simulations to mimic the observed spectra, and then simulated spectra are generated following a Poisson distribution. Based on the EP-WXT light curve and best-fitting spectral model of EP240315a, such a procedure is repeated 10,000 times to produce 10,000 simulated ``observed'' light curve at different redshifts by taking Poisson distribution into account. We then determine the EP-WXT signal-to-noise ratio at different redshifts by calculating the median SNR value of simulated light curves. 
The evolution of the simulated EP-WXT signal-to-noise ratio as a function of redshift is illustrated in Extended Data Fig. \ref{fig:simulate_lc_wxtfxt}a. It is demonstrated that the EP240315a-like event can be well detected at a redshift of 7.5 by EP-WXT with an SNR $\ge7$ and even higher redshift with a lower SNR (Extended Data Fig. \ref{fig:simulate_lc_wxtfxt}b).

\section*{Data Availability}

The processed data are presented in the tables and figures of the paper, which are available upon reasonable request. The authors point out that some data used in the paper are publicly available, whether through the UK Swift Science Data Centre website, or GCN circulars.

\section*{Code Availability}
Upon reasonable requests, the code (mostly in Python) used to produce the results and figures will be provided.

\bigskip
\bigskip
\bigskip


\bigskip
\bigskip
\bigskip

\begin{addendum}

\item[Acknowledgments] This work is based on the data obtained with Einstein Probe, a space mission supported by Strategic Priority Program on Space Science of Chinese Academy of Sciences, in collaboration with ESA, MPE and CNES (Grant No. XDA15310000),
the Strategic Priority Research Program of the Chinese Academy of Sciences (Grant No. XDB0550200), and the National Key R\&D Program of China (2022YFF0711500). We acknowledge the support by the National Natural Science Foundation of China (Grant Nos. 12321003, 12103065, 12333004, 12373040, 12021003), the China Manned Space Project (Grant Nos. CMS-CSST-2021-A13, CMS-CSST-2021-B11), and the Youth Innovation Promotion Association of the Chinese Academy of Sciences. We acknowledge the data resources and technical support provided by the China National Astronomical Data Center, the Astronomical Science Data Center of the Chinese Academy of Sciences, and the Chinese Virtual Observatory. We acknowledge the observational data taken at NOT (programs 68-811, PI Xu, and 68-020, PI Malesani), VLT (program 110.24CF, PIs: Tanvir, Vergani, Malesani), TNG: (program A47TAC 42, PI: Melandri), and LBT (program IT-2023B-020, PI Maiorano). The work of D.S.S., D.D.F, A.V.R., A.L.L., A.E.T., M.V.U., A.G.D. and A.A.K. was carried out in the framework of the basic funding programme of the Ioffe Institute no. FFUG-2024-0002; A.E.T also acknowledges financial support from Accordo ASI e INAF HERMES 2022-25-HH.0. P.G.J.~has received funding from the European Research Council (ERC) under the European Union’s Horizon 2020 research and innovation programme (Grant agreement No.~101095973). R. R. and E.T.~acknowledge support from the European Research Council (ERC) under the European Union’s Horizon 2020 research and innovation programme (Grant agreement No.~101002761). L.P., G.B., G.G., and A.L.T. acknowledge useful discussions with Lauren Rhodes regarding the analysis and interpretation of the e-MERLIN dataset. L.P., G.B., G.G. and A.L.T. acknowledge support from the European Union Horizon 2020 programme under the AHEAD2020 project (grant agreement number 871158). L.P., G.G. and A.L.T. also acknowledge support from the ASI (Italian Space Agency) through Contract No. 2019-27-HH.0. 
P.O.'B. and N.R.T. acknowledge support from UK/STFC grant ST/W000857/1. J.D. acknowledge support from NASA contract NAS5-0136.

\item[Author Contributions] W.Y. initiated the Einstein Probe mission and leads the project as the Principal Investigator. Y.L., H.S., H.G., X.-F.W., B.Z. initiated the study. Y.L, X.-F.W., B.Z., H.G., H.S., D.X. coordinated the scientific investigations of the event. Y.L., H.S., W.-J.Z., D.-Y.L., J.Y. and Y.-H.I.Y. processed and analysed the WXT data. Q.-Y.W., C.-K.L., Y.C. processed and analysed the FXT data. J.-W.H. analysed the BAT light curves and simulated the detectability of WXT. J.-D.L., J.A., A.L., Y.-N.M., H.G., D.X., J.-J.G. performed the multi-wavelength afterglow modelling. H.G., X.-F.W., B.Z. led the theoretical investigation of the event. 
J.D. performed the GRB search in Swift/BAT data, as well as developing the likelihood analysis for the spectral fit, light curve, and localisation. G.R. performed the $T_{90}$ fit. Comments and contributions were provided by the rest of the Swift BAT GUANO team (J.A.K., T.P., S.R., A.T.)
D.S.S. and D.D.F. performed GRB search in the Konus-Wind data,
the high-energy spectral analysis and upper limit calculations with
the contributions of the Konus-Wind team (A.V.R., A.L.L., A.E.T., M.V.U., A.G.D. and A.A.K.).
B.-B.Z., J.Y. and Y.-H.I.Y. performed GRB search in Fermi/GBM data. J.Y., A.L. contributed to the Amati relation. 
D.X., P.G.J., N.R.T., S.D.V, D.B.M., A.J.L., A.dU.P., S.P.L., A.M-C., J.Q.V., A.R., B.S., M.A.P.T., D.M.S., M.E.R., Y.D.H., J.P.U.F. contributed to the optical and near-infrared data taking and analysis and provided comments to the manuscript.
R.R., E.T., D.D. and J.K.L. acquired and analyzed the ATCA data and provided comments to the manuscript. 
L.P., G.B., A.L.T. and G.G. contributed to the e-MERLIN radio data acquisition and provided comments on the manuscript. G.B. performed the analysis of the e-MERLIN radio data. Z.-X.L., C.Z., X.-J.S., S.L.S., X.-F.Z., Y.-H.Z., Z.-M.C. F.-S.C. and W.Y. contributed to the development of the WXT instrument. C.Z., Z.-X.L., H.-Q.C., D.-H.Z. and Y.L. contributed to the calibration of WXT data.
Y.L., H.-Q.C., C.J., W.-D.Z., D.-Y.L., J.-W.H., H.-Y.L., H.S., H.-W.P. and M.-J.L. contributed to the development of WXT data analysis software.
Y.C., S.-M.J., W.-W.C., C.-K.L., D.-W.H., J.W., W.L., Y.-J.Y., Y.-S.W., H.-S.Z., J.G., J.Z., X.-F.Z., J.-J.X., J.M., L.-D.L., H.W., X.-T.Y., T.-X.C., J.H., Z.-J.Z., Z.-L.Z., M.-S.L., Y.-X.Z., D.-J.H., L.-M.S., F.-J.L., C.-Z.L., Q.-J.T. and H.-L.C. contributed to the development of the FXT instrument. 
S.-M.J., H.-S.Z., C.-K.L., J.Z. and J.G. contributed to the development of FXT data analysis software.
Y.L., H.G., H.S., B.Z., Z.-P.Z, J.-W.H., P.O.'B., Y.J. D.X. contributed to the interpretation of the observations and the writing of the manuscript with contributions from all authors.

\item[Competing Interests] The authors declare that they have no competing financial interests.

\end{addendum}

\clearpage
\setcounter{figure}{0}
\setcounter{table}{0}

\captionsetup[figure]{labelfont={bf}, labelformat={default}, labelsep=period, name={Extended Data Fig.}}
\captionsetup[table]{labelfont={bf}, labelformat={default}, labelsep=period, name={Extended Data Table}}


\begin{figure}
\centering
\includegraphics[width=15cm]{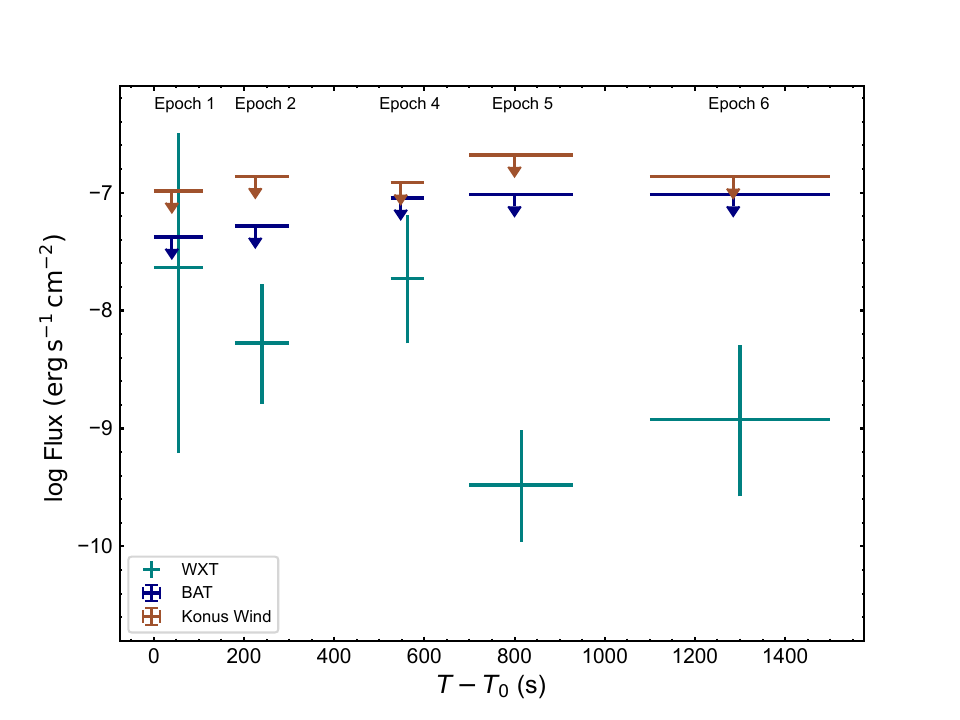}	
\caption{\noindent\textbf{The upper limits (15--350 keV) of BAT and Konus-Wind in the epochs without gamma-ray detection.} The extrapolated fluxes (with $1\sigma$ uncertainty) from the WXT spectra are also shown for comparison and well below the upper limits in these epochs.}
\label{fig:fluxlimit}
\end{figure}

\clearpage

\begin{figure}
\centering
\includegraphics[width=12cm]{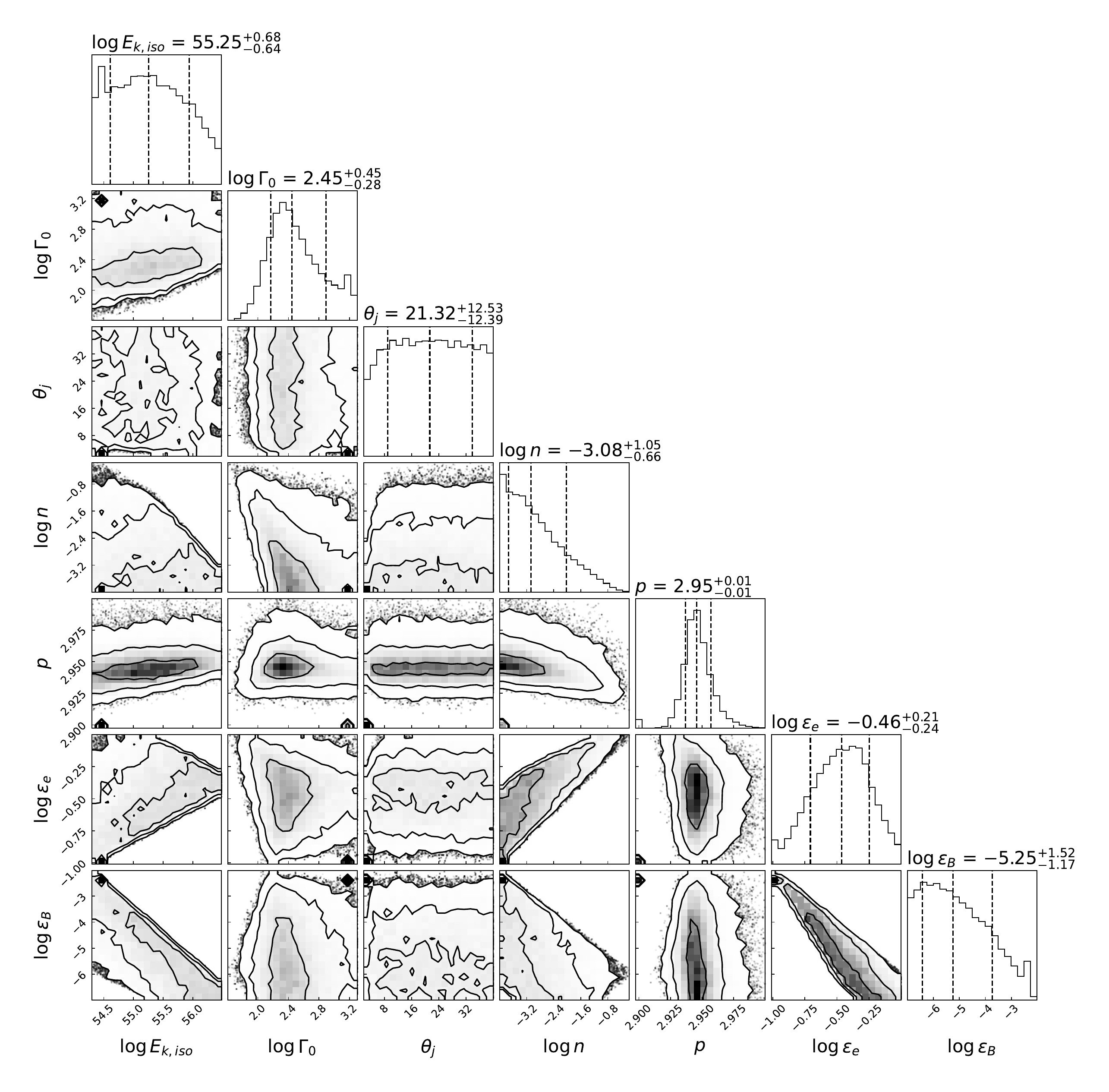}
\caption{\noindent\textbf{The Multi-wavelength afterglow model.} The corner plot of the posterior probability distributions of the parameters for the standard GRB afterglow model.
In the 1D histogram, the three dashed lines signify the median of the sample distribution along with the range of $1\sigma$. Meanwhile, the contour curves within the 2D histogram, as delineated by the ``corner'' package, correspond to the confidence levels of $1\sigma, 2\sigma, 3\sigma$ of a two-dimensional Gaussian distribution.
}
\label{fig:MCMC1}
\end{figure}

\clearpage

\begin{figure}
\centering
\includegraphics[width=12cm]{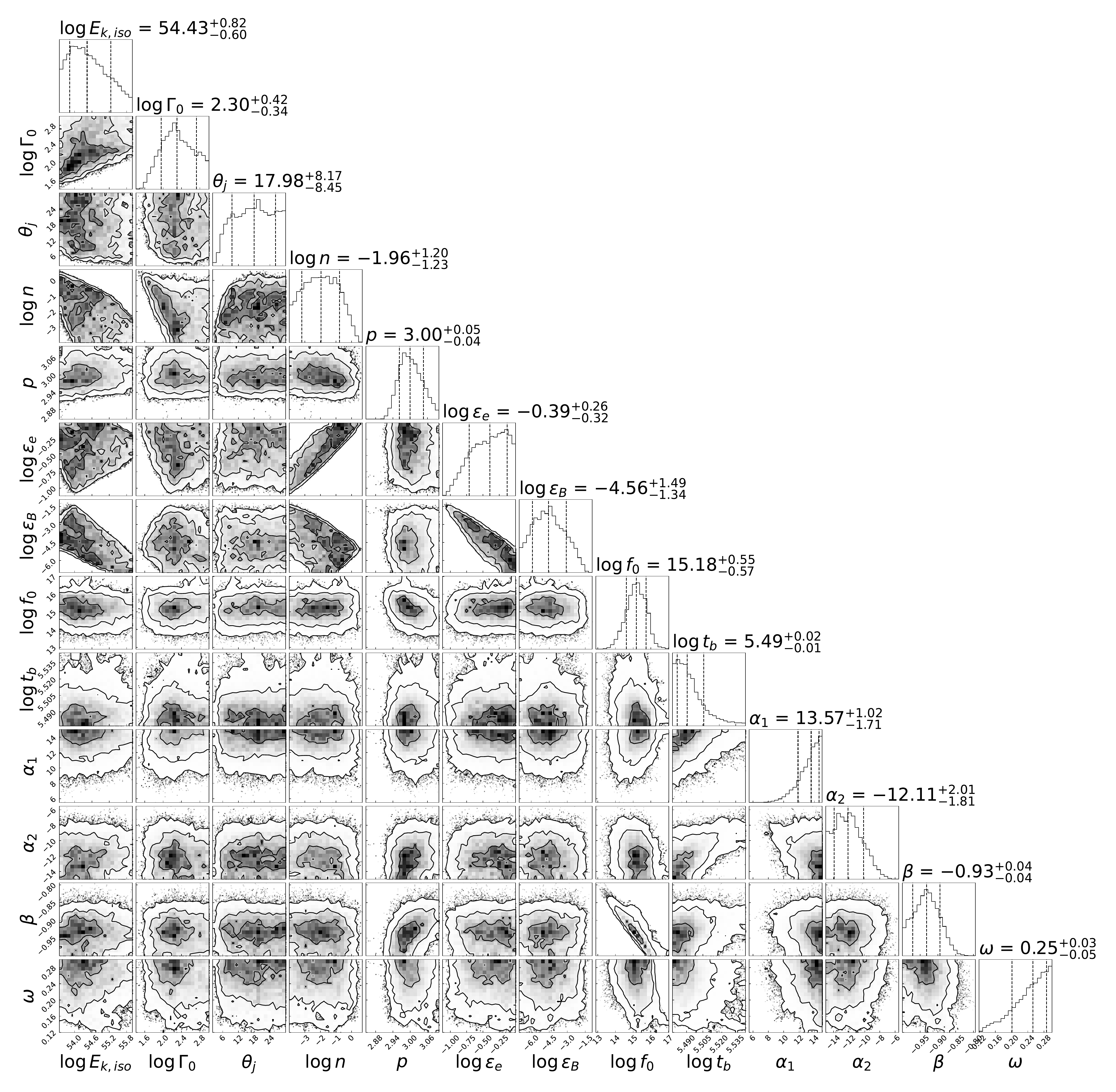}
\caption{\noindent\textbf{The Multi-wavelength afterglow model with the re-brightening signature incorporated.} Same as Extended Data Fig. \ref{fig:MCMC1}. The corner plot of the posterior probability distributions of the parameters for the standard GRB afterglow model, where a late re-brightening signature in X-ray and optical being incorporated.}
\label{fig:MCMC2}
\end{figure}

\clearpage

\begin{figure}
\centering
\begin{tabular}{c}
\begin{overpic}[width=0.47\textwidth]{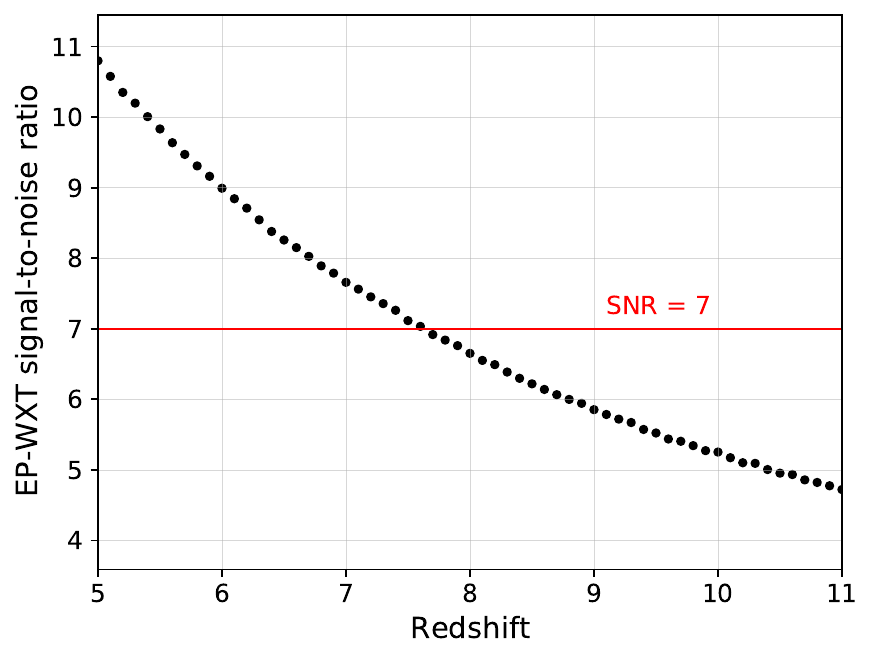}\put(0, 75){\bf a}\end{overpic}
\begin{overpic}[width=0.52\textwidth]{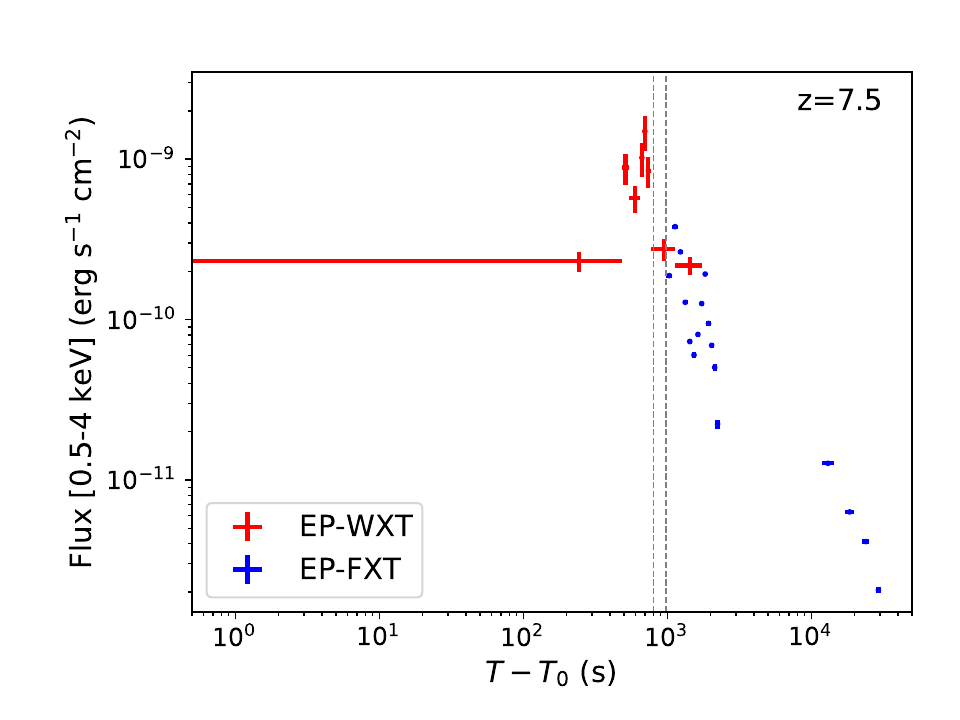}\put(6, 67){\bf b}\end{overpic} \\
\end{tabular}
\caption{\noindent\textbf{Simulations of EP240315a-like events by EP-WXT at different redshifts.} \textbf{a}, The simulated EP-WXT signal-to-noise ratio of a EP240315a-like event as the function of redshift. \textbf{b}, An example of the simulated EP observations of EP240315a-like events at a redshift of 7.5. The red and blue data points represent the simulated WXT and FXT light curves, respectively. The left and right dashed vertical lines mark the time to trigger FXT follow-up and the start time of FXT observations after taking into account the slew maneuver, respectively.}
\label{fig:simulate_lc_wxtfxt}
\end{figure}

\clearpage

\begin{table*}
\centering
\small
\caption{Log of the EP-FXT follow-up observations.}\label{tab:fxt_obsinfo}
\begin{tabular}{ccccc}
\hline
ObsID & START$\;$TIME & END$\;$TIME&DURATION&EXPOSURE\\
 & (UTC) &(UTC)&(second)&(second)\\
 \hline
08500000025 &2024-03-17T14:14:23& 2024-03-17T16:05:26 &6,663 &3,758\\
08500000026 &2024-03-18T16:04:13& 2024-03-18T19:23:05 &11,932 &5,649\\
08500000027 &2024-03-19T04:10:50& 2024-03-19T08:13:46 &14,575 &8,993\\
08500000038 &2024-03-19T20:25:11& 2024-03-19T22:40:47 &8,136 &4,383\\
08500000039 &2024-03-20T05:02:48& 2024-03-20T08:18:50 &11,762 &6,184\\
08500000042 &2024-03-20T16:01:57& 2024-03-20T19:33:13 &12,676 &6,540\\
08500000041 &2024-03-21T00:02:34& 2024-03-21T03:34:56 &12,741 &7,165\\
08500000043 &2024-03-21T15:35:19& 2024-03-21T18:02:02 &8,802 &5,878\\
06800000001 &2024-03-23T15:45:42& 2024-03-23T19:48:53 &14,590 &8,642\\
08503014656 &2024-03-25T11:33:22& 2024-03-26T07:12:40 &70,758 &37,098\\
\hline
\end{tabular}
\end{table*}

\clearpage

\begin{scriptsize}
\begin{longtable}{c c c c c}
\caption{The photometric results of our observations, combined with the data obtained from GCN Circular. $\Delta T$ is the expose medium time after the EP trigger. Magnitudes are not corrected for Galactic Extinction, which is E(B-V) = 0.05\cite{Schlafly2011ApJ}.} \label{tab:phot} \\
\hline \multicolumn{1}{c}{$\Delta T$(days)} & \multicolumn{1}{c}{Telescope/Observatory} & \multicolumn{1}{c}{Band} & \multicolumn{1}{c}{Mag(AB)} & \multicolumn{1}{c}{Reference} \\ \hline 
\endfirsthead
\multicolumn{5}{c}%
{{\bfseries \tablename\ \thetable{} -- continued from previous page}} \\
\hline \multicolumn{1}{c}{$\Delta T$(days)} & \multicolumn{1}{c}{Telescope/Observatory} & \multicolumn{1}{c}{Band} & \multicolumn{1}{c}{Mag(AB)} & \multicolumn{1}{c}{Reference} \\ \hline 
\endhead
\hline \multicolumn{5}{r}{{Continued on next page}} \\
\endfoot
\hline
\multicolumn{5}{l}{{References: (1) this work, (2) TNS report, (3) GCN report}}
\endlastfoot
0.852 & HMT / Xingming Obs. & white & $>$ 20.9 & (1)\\
1.086 & GTC & r & 21.74$\pm$0.06 & (1)\\
1.104 & GTC & i & 21.14$\pm$0.04 & (1)\\
2.041 & GTC & g & $>$ 25.3 & (1)\\
2.041 & GTC & r & 23.99$\pm$0.08 & (1)\\
2.041 & GTC & i & 22.32$\pm$0.03 & (1)\\
2.041 & GTC & z & 22.05$\pm$0.03 & (1)\\
3.139 & GTC & J & 21.87$\pm$0.15 & (1)\\
3.097 & GTC & K & 21.09$\pm$0.1 & (1)\\
4.045 & GTC & K & 21.47$\pm$0.12 & (1)\\
16.067 & GTC & z & 24.55$\pm$0.41 & (1)\\

1.195 & VLT & r & 22.68$\pm$0.05 & (1)\\
1.198 & VLT & g &  $>$ 25  & (1)\\
1.201 & VLT & z & 21.06$\pm$0.06 & (1)\\
7.233 & VLT & z & 23.92$\pm$0.11 & (1)\\
16.183 & VLT & z & 25.44$\pm$0.45 & (1)\\

1.756 & Xinglong 2.16m & R & $>$ 20.5 & (1)\\

2.015 & TNG & H & 21.15$\pm$0.2 & (1)\\
4.052 & TNG & H & 21.56$\pm$0.19 & (1)\\

2.061 & NOT & z & 22.02 $\pm$ 0.11 & (1)\\
3.194 & NOT & z & 22.23$\pm$0.11 & (1)\\
5.160 & NOT & z & 23.1 $\pm$ 0.3 & (1)\\

4.3 & LBT & J & 22.49$\pm$0.08 & (1)\\
6.4 & LBT & H & 23.29$\pm$0.22 & (1)\\

3.951 & ATCA & 5.5 GHz &  74 $\pm$ 21 $\mu$Jy & (1)\\
3.951 & ATCA & 9.0 GHz &  82 $\pm$ 20 $\mu$Jy & (1)\\
10.524 & ATCA & 5.5 GHz & 97 $\pm$ 14 $\mu$Jy & (1)\\
10.524 & ATCA & 9.0 GHz & 158 $\pm$ 12 $\mu$Jy & (1)\\
19.576 & ATCA & 5.5 GHz & 140 $\pm$ 27 $\mu$Jy & (1)\\
19.576 & ATCA & 9.0 GHz & 328 $\pm$ 31 $\mu$Jy & (1)\\
5.066 & e-MERLIN & 5 GHz & $<$ 75 $\mu$Jy  & (1) \\
8.550 & e-MERLIN & 5 GHz & $67 \pm 13$ $\mu$Jy & (1) \\
12.034 & e-MERLIN & 5 GHz & $<$ 105 $\mu$Jy  & (1) \\

1.490 & Pan-STARRS & i & 21.91 $\pm$ 0.2 & (2)\\

0.965 & PRIME & J & 20.64 $\pm$ 0.3 & (3)\cite{Guiffreda2024GCN35956}\\

1.012 & LT & i & $>$ 20.02 & (3)\cite{Srivastav2024GCN35933}\\
1.277 & GROND & J & 20.5 $\pm$ 0.2 & (3)\cite{Rau2024GCN5937}\\
2.884 & MeerKAT & 3 GHz & $\sim$ 30 $\pm$ 8.5 $\mu$Jy/beam & (3)\cite{Carotenuto2024GCN35961}\\
4.356 & P200 / Palomar Obs. & J & $\sim$ 22.5 & (3)\cite{Earley2024GCN36008}\\
\end{longtable}
\end{scriptsize}

\clearpage

\begin{table*}
\centering
\small
\caption{Konus-Wind spectral fits. All errors represent the $1\sigma$ uncertainties.}
\label{tab:kw_spec}
\begin{tabular}{cccccc}
\hline
Spectrum & Time Interval    & Model &     $\alpha_{\rm KW}$       &   $E_\mathrm{peak}$   &      Flux                   \\
         &     (s)          &       &                    &     (keV)      & ($\rm 10^{-6} \,erg\,cm^{-2}\,s^{-1}$)\\
\hline
       & 372.516--387.236 &  CPL  & $-1.04_{-0.46}^{+6.04}$ &  $309_{-191}^{+851}$ &  $0.21_{-0.10}^{+0.17}$ \\
Peak   & 387.236--396.068 &  CPL  & $-0.86_{-0.21}^{+0.24}$ &  $428_{-108}^{+168}$ &  $0.75_{-0.13}^{+0.16}$\\
       & 396.068--407.844 &   PL  & $-1.53_{-0.13}^{+0.12}$ &        --            &  $1.02_{-0.30}^{+0.40}$\\
       & 407.844--416.676 &  CPL  & $-1.03_{-0.42}^{+6.03}$ &  $197_{-92}^{+170}$  &  $0.38_{-0.14}^{+0.13}$ \\
Total  & 372.516--416.676 &  CPL  & $-1.19_{-0.14}^{+0.16}$ &  $441_{-117}^{+207}$ &  $0.43_{-0.06}^{+0.09}$ \\
\hline
\end{tabular}
\end{table*}

\clearpage

\begin{table*}
\centering
\scriptsize
\begin{threeparttable}
\caption{Spectral properties of late X-ray observations of EP240315a.}
\label{tab:xaftgl_spec}
\begin{tabular}{cccccc}
\toprule
Instruments & Intrinsic Absorption & Spectral Index & Time Intervals & Absorbed Flux \tnote{1,2} & CSTAT/(d.o.f) \\
 & ($ \rm{cm^{-2}}$) & & (ks) & ($\rm erg\,cm^{-2}\,s^{-1}$) &  \\
\hline
EP-WXT & $2.5^{+13.8}_{-2.5} \times 10^{22}$ & $-1.9^{+0.8}_{-0.8}$ & 5.7--7.6 & $4.3^{+1.2}_{-1.2} \times 10^{-11}$ & 37.84/15 \\
 & - & -  & 10.2--13.4 & $1.3^{+0.5}_{-0.4} \times 10^{-11}$ &  \\
 & - & -  & 16.1--19.2 & $<9.5 \times 10^{-12}$ &  \\
 & -  & -  & 27.6--30.8 & $<1.7 \times 10^{-11}$ &  \\
\hline
\multirow{9}{*}{EP-FXT \tnote{3}} & \multirow{9}{*}{$1.0^{+1.3}_{-1.0}\times 10^{23}$} & \multirow{9}{*}{$-1.9^{+0.5}_{-0.5}$}& 151.4--158.1 & $5.0^{+1.4}_{-1.2} \times 10^{-14}$ & \multirow{9}{*}{394.61/443}\\
 & & & 244.4--256.3 & $2.1^{+0.8}_{-0.6} \times 10^{-14}$ &  \\
 & & & 288.0--302.6 & $3.3^{+0.8}_{-0.7} \times 10^{-14}$ &  \\
 & & & 346.5--354.6 & $3.3^{+1.1}_{-0.9} \times 10^{-14}$ &  \\
 & & & 377.5--389.3 & $2.6^{+0.8}_{-0.7} \times 10^{-14}$ &  \\
 & & & 417.1--429.7 & $1.6^{+0.7}_{-0.6} \times 10^{-14}$ &  \\
 & & & 445.9--458.7 & $9.4^{+5.5}_{-4.7} \times 10^{-15}$ &  \\
 & & & 501.9--510.7 & $1.3^{+0.7}_{-0.6} \times 10^{-14}$ &  \\
 & & & 675.3--689.9 & $5.8^{+5.5}_{-4.7} \times 10^{-15}$ &  \\
\hline
\multirow{2}{*}{Chandra-ACIS \tnote{4}} & \multirow{2}{*}{same as EP-FXT} & \multirow{2}{*}{-}& 258.2--270.5 & $4.4^{+0.8}_{-0.7} \times 10^{-14}$ &  
\multirow{2}{*}{-} \\
 & & & 894.3--915.3 & $<4.1 \times 10^{-15}$ &  \\
\bottomrule
\end{tabular}
\begin{tablenotes}
\footnotesize
\item[1] The absorbed flux is derived in 0.5--4 keV for EP-WXT, EP-FXT and Chandra-ACIS.
\item[2] Systematic errors are not included.
\item[3] The spectra of the FXT observations are fitted simultaneously with \textit{XSPEC} model {\it tbabs*ztbabs*powerlaw} with the redshift fixed to $z=4.859$. 
\item[4] Since the numbers of photons are limited in the two Chandra observations, the flux and upper limit are calculated under the same model parameters as those of EP-FXT.
\end{tablenotes}
\end{threeparttable}
\end{table*}

\clearpage

\begin{table}
\caption{The posterior probability distribution of model parameters and the BIC values of two fitting approaches.}
\label{parameters}
\scriptsize
\centering
\begin{tabular}{ l c c c }
\hline
Parameters & & Standard Afterglow & Standard Afterglow with re-brightening\\
\hline
Initial Lorentz factor & $\Gamma_{0}$ & $282_{-134}^{+512}$ & $199_{-108}^{+325}$ \\
Isotropic kinetic energy & $E_{\rm k,iso}$ $\rm (erg)$ & $1.8_{-1.4}^{+6.7}\times10^{55}$ & $2.7_{-2.0}^{+15.1}\times10^{54}$\\
Half-opening angle & $\theta_{\rm j}$ $\rm (deg)$ & $21.3_{-12.4}^{+12.5}$ & $18.0_{-8.5}^{+8.2}$\\
Density of ISM & $n$ $\rm (cm^{-3})$ & $8.3_{-6.5}^{+85.0}\times10^{-4}$ & $1.1_{-1.0}^{+16.3}\times10^{-2}$\\
Distribution index of electrons & $p$ & $2.95_{-0.01}^{+0.01}$ & $3.00_{-0.04}^{+0.05}$\\
Fraction of shock energy into electrons & $\epsilon_e$ & $0.35_{-0.15}^{+0.21}$ & $0.41_{-0.22}^{+0.33}$\\
Fraction of shock energy into magnetic field & $\epsilon_B$ & $6_{-5}^{+180}\times10^{-6}$ & $3_{-3}^{+80}\times10^{-5}$\\
Re-brightening break time & $t_{\rm b}$ $\rm (s)$ & - & $3.1_{-0.1}^{+0.2}\times10^{5}$\\
Re-brightening rising temporal slope & $\alpha_1$ & - & $13.6_{-1.2}^{+1.0}$\\
Re-brightening descent temporal slope & $\alpha_2$ & - & $-12.1_{-1.8}^{+2.0}$\\
Sharpness of re-brightening peak & $\omega$ & - & $0.25_{-0.05}^{+0.03}$\\
Re-brightening spectral index & $\beta$ & - & $-0.93_{-0.04}^{+0.04}$\\
Amplitude parameter of re-brightening & $F_{0}$ $(\rm mJy)$ & - & $1.5_{-1.1}^{+3.9}\times10^{15}$\\
\hline
BIC & & $170.87$ & $114.57$\\
\hline
\end{tabular}
\end{table}

\end{document}